\documentclass[twocolumn]{aastex62}
\usepackage{multirow}

\shorttitle{WorldWide Telescope}
\shortauthors{Rosenfield et al.}

\begin{document}

\title{AAS WorldWide Telescope: Seamless, Cross-Platform Data Visualization Engine for Astronomy Research, Education, and Democratizing Data}

\correspondingauthor{Philip Rosenfield}
\email{philip.rosenfield@aas.org}

\author[0000-0001-9306-6049]{Philip Rosenfield}
\affil{American Astronomical Society \\
1667 K St NW Suite 800 \\
Washington, DC 20006, USA}

\author[0000-0003-2500-8984]{Jonathan Fay}
\affiliation{American Astronomical Society \\
1667 K St NW Suite 800 \\
Washington, DC 20006, USA}

\author[0000-0001-8246-5001]{Ronald K Gilchrist}
\affiliation{American Astronomical Society \\
1667 K St NW Suite 800 \\
Washington, DC 20006, USA}

\author[0000-0002-7456-1826]{Chenzhou Cui}
\affiliation{National Astronomical Observatories, Chinese Academy of Sciences\\
20A Datun Road, Chaoyang District\\
Beijing, 100012, China}

\author[0000-0002-8026-2291]{A. David Weigel}
\affiliation{Christenberry Planetarium, Samford University \\
800 Lakeshore Drive \\
Birmingham, AL 35229, USA}

\author[0000-0002-8642-1329]{Thomas Robitaille}
\affiliation{Aperio Software Ltd.\\
Headingley Enterprise and Arts Centre, Bennett Road \\
Leeds, LS6 3HN, United Kingdom}

\author[0000-0002-4679-5692]{Oderah Justin Otor}
\affiliation{American Astronomical Society \\
1667 K St NW Suite 800 \\
Washington, DC 20006, USA}

\author[0000-0003-1312-0477]{Alyssa Goodman}
\affiliation{Harvard Smithsonian Center for Astrophysics \\
60 Garden St. \\
Cambridge, MA 02138}

\begin{abstract}

The American Astronomical Society's WorldWide Telescope (WWT) project enables terabytes of astronomical images, data, and stories to be viewed and shared among researchers, exhibited in science museums, projected into full-dome immersive planetariums and virtual reality headsets, and taught in classrooms from middle school to college levels. We review the WWT ecosystem, how WWT has been used in the astronomical community, and comment on future directions.

\end{abstract}


\section{Introduction} \label{sec:intro}

WorldWide Telescope\footnote{http://worldwidetelescope.org} (WWT) was first designed as a tool for astrophysical data exploration and discovery in the era of large and disparate data \citep[][]{Gray2004}. WWT was an early attempt to answer an emerging question, ``how will research techniques change in the era of multi-wavelength, multi-epoch, large datasets hosted throughout the Web?'' Launched in 2008 by Microsoft Research, WWT was open-sourced (under the MIT license) seven years later with copyrights transferred to the .NET Foundation and project direction and management taken up by the American Astronomical Society (AAS). Recent and near future AAS WWT development is focused on implementing operating system-agnostic features to better enable astronomers to use WWT either individually or through archival data centers.

\begin{figure*}
\includegraphics[width=0.48\textwidth]{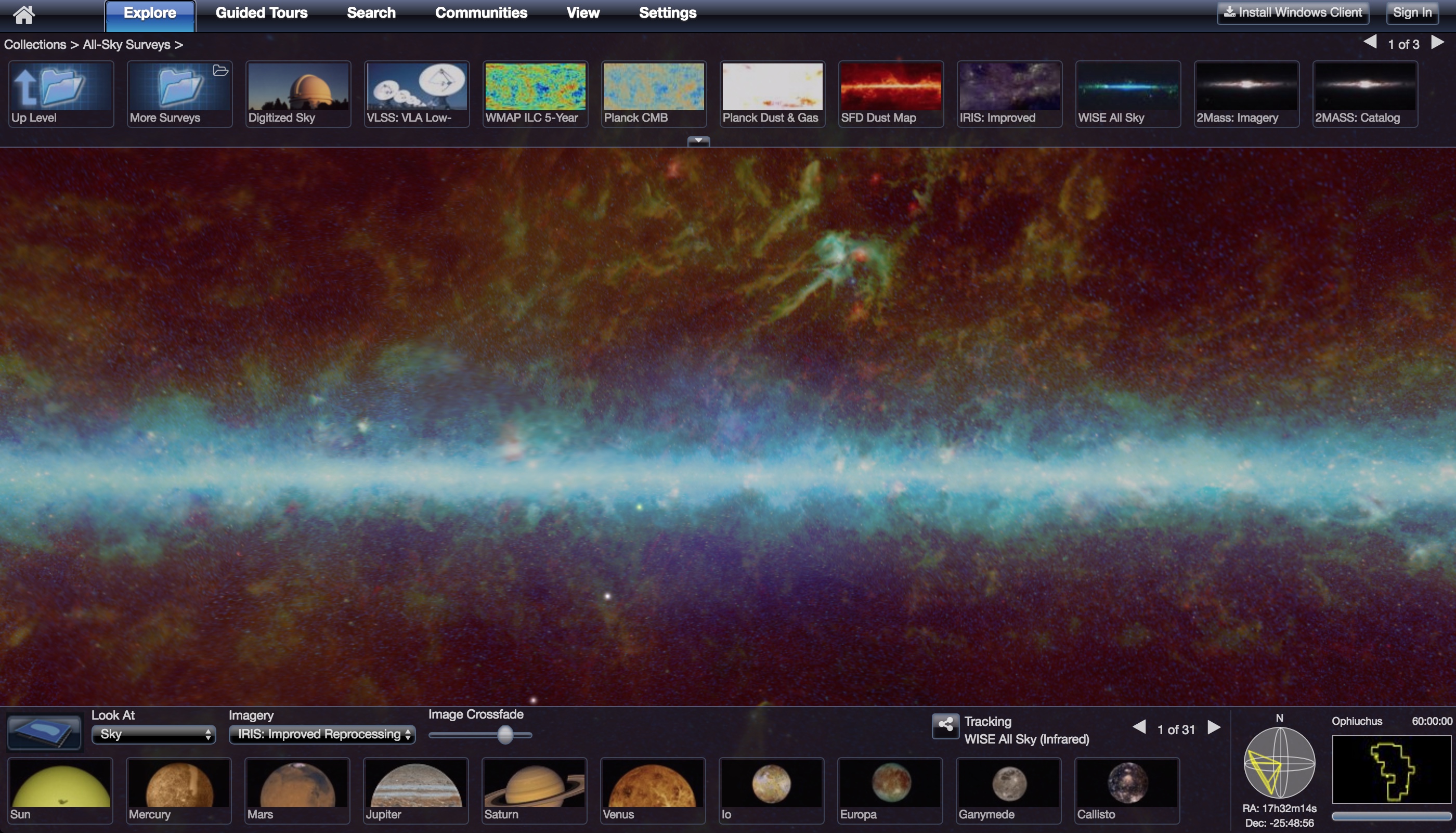}
\includegraphics[width=0.48\textwidth]{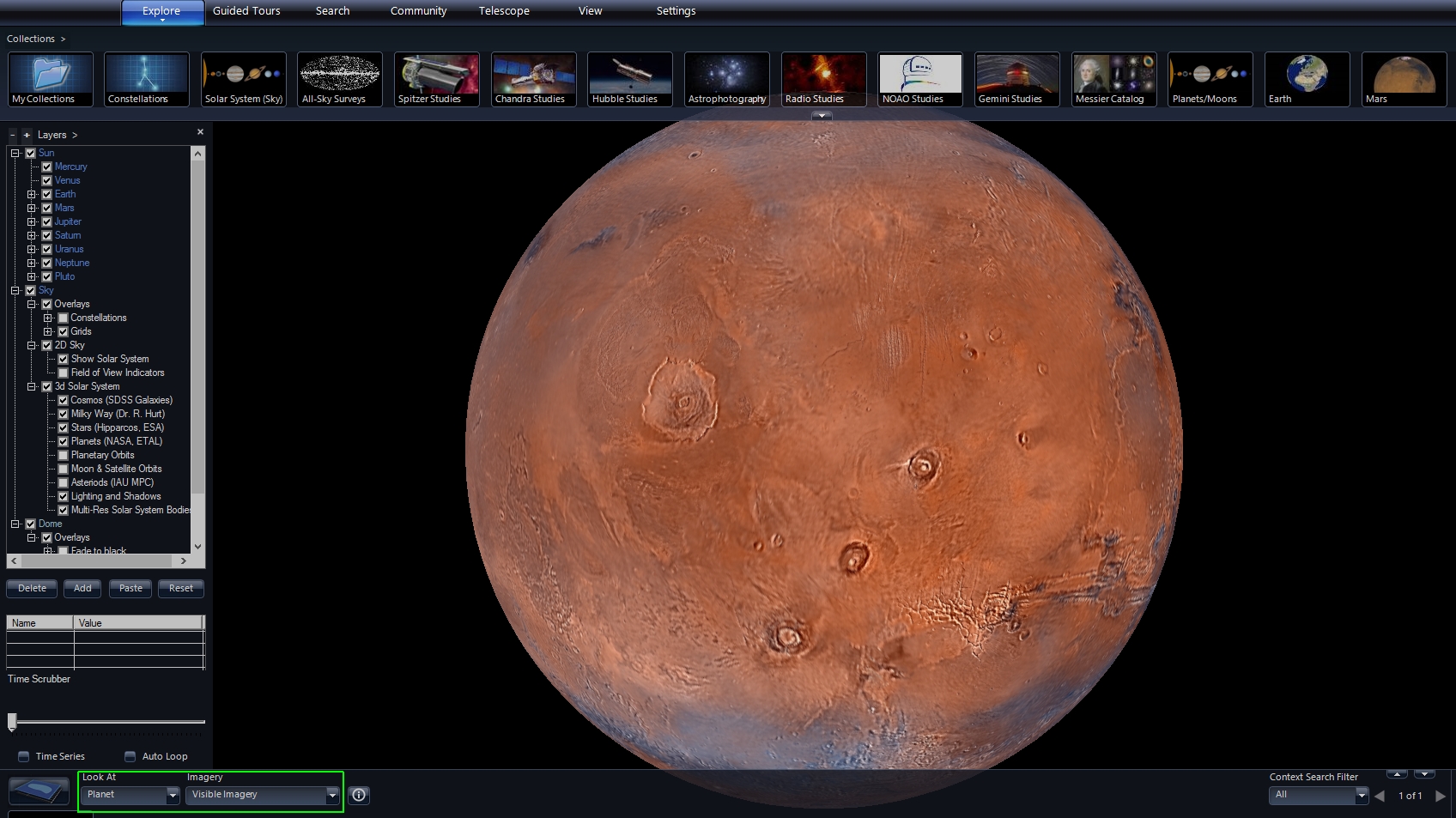} \\
\includegraphics[width=0.48\textwidth]{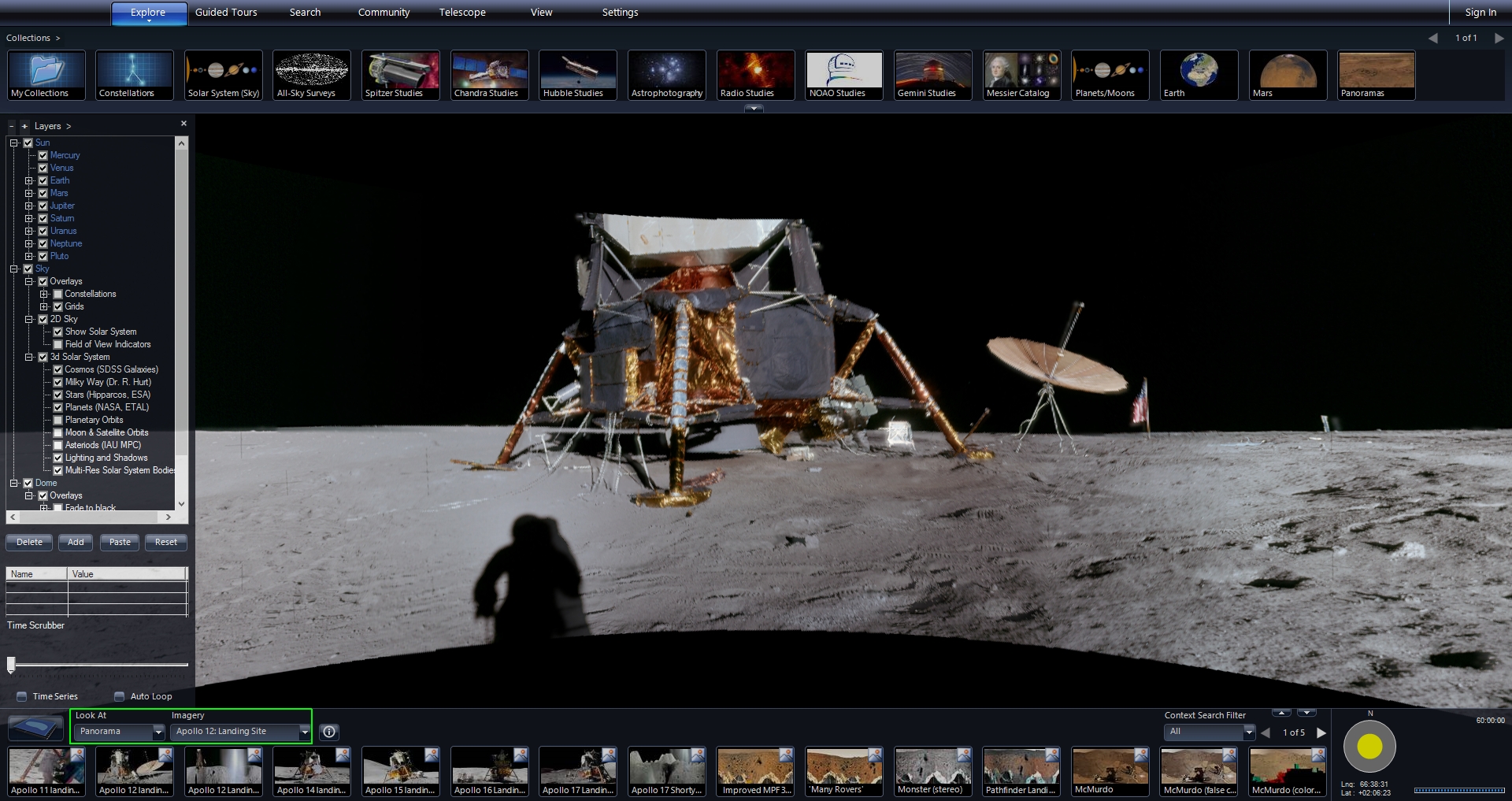}
\includegraphics[width=0.48\textwidth]{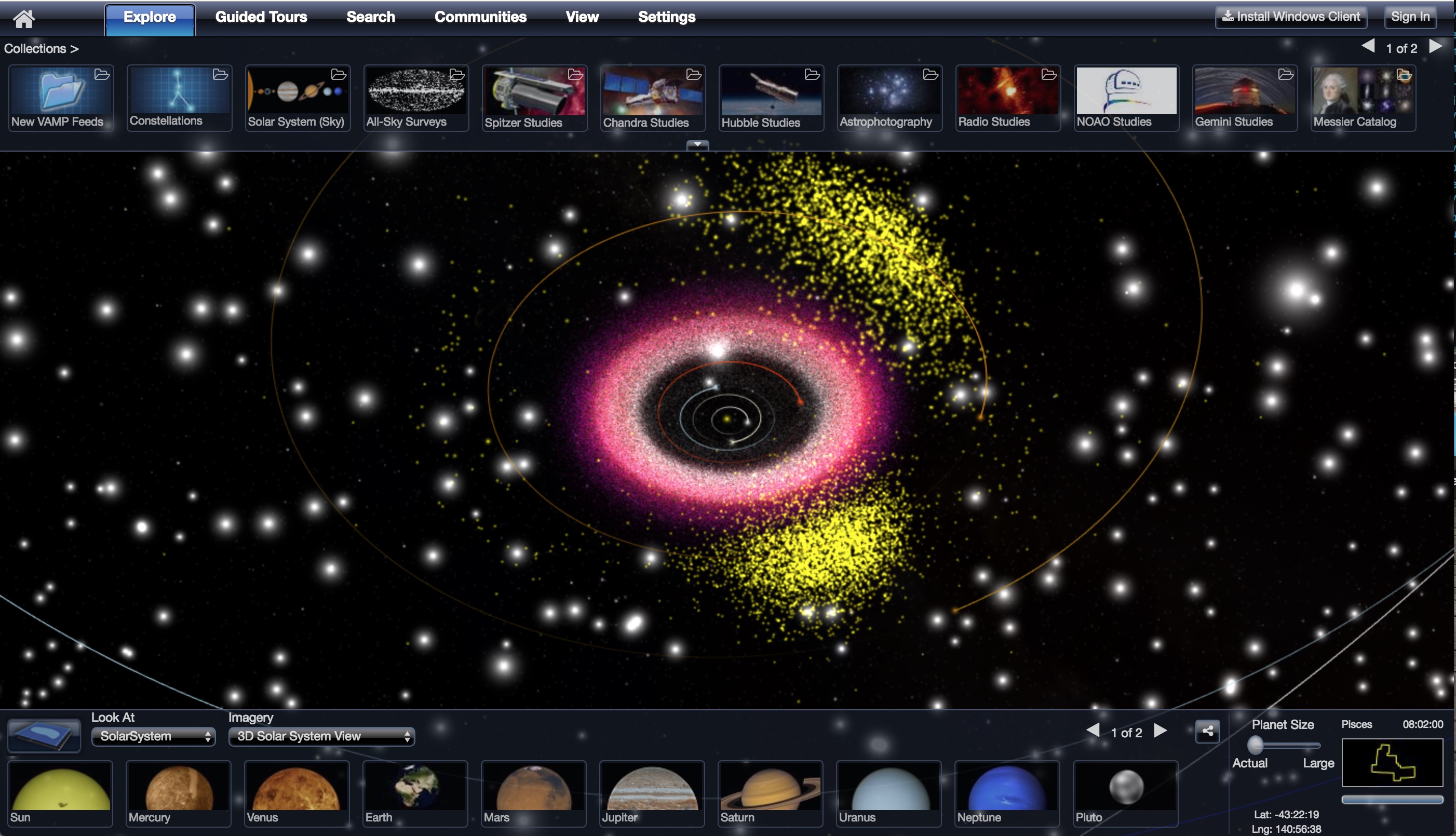}
 \caption{Screen captures of the WorldWide Telescope User Interface. Clockwise from top left: ``Sky'' mode showing IRIS: Improved Reprocessing of Infrared Astronomy Satellite (IRAS) Survey (12, 60, and 100 $\mu$m (Red is IRIS100, Green IRIS60, Blue IRIS12) cross faded against 100 $\mu$m David J. Schlegel, Douglas P. Finkbeiner and Marc Davis (SFD) Dust Map (Data provided by IRAS and the Cosmic Background Explorer, Princeton University and University of California, Berkeley. Original IRAS data: NASA/JPL IPAC, IRIS Reprocessing: Canadian Institute for Theoretical Astrophysics/Institut d'Astrophysique Spatiale.  TOAST-formatted mosaics were obtained using facilities of NASA’s SkyView Virtual Telescope.); ``Planet'' mode showing Mars visible imagery (NASA/USGS/Malin Space Science Systems/JPL); ``3D Solar System'' mode (which includes HIPPARCHOS and COSMOS data) showing planets and their orbital paths, Oort cloud objects (white), Jupiter Trojans (yellow), and main belt asteroids (purple to white, colored by zone. Minor Planet ephemerides from the International Astronomical Union's Minor Planet Center); and ``Panorama'' mode showing an image taken by Apollo 12 Lunar Module Pilot Alan Bean in 1969 and assembled at the NASA Ames Research Center in 2007. (Images courtesy of NASA and the Lunar and Planetary Institute). \label{fig:wwtui}}
\end{figure*}

WWT is a scientific data visualization platform (see Figure \ref{fig:wwtui}). It acts as a virtual sky, allowing users to explore all-sky surveys across the electromagnetic spectrum, to overlay data from NASA's Great Observatories, and import their own imagery and tabular data. In WWT's 3D environment, users can explore planetary surfaces and elevation maps; orbital paths and ephemerides of major and minor solar system bodies as well as asteroids; zoom out from the Solar System to view the positions and approximate colors of stars from the HIPPARCOS catalog \citep{Perryman1997} and galaxies of the Cosmic Evolution Survey dataset \citep[COSMOS;][]{Scoville2007}.

WWT's unique contextual narrative layer \citep{Wong2008} has set it apart from the myriad of other available data visualization systems.  Users can record or view recorded paths though the virtual environments and add narration, text, and imagery. With these ``tours'' users can share their stories, outreach specialists can create programming, and researchers can distill their discoveries to other researchers or to the public.

These features make WWT a seamless data visualization program with an engaging learning environment. This combination has enabled WWT to deliver astrophysical data hosted around the world not only to researchers, but to classrooms, planetariums, and educational outreach programs. Therefore, WWT is more than a tool to aid astronomical research and discovery, it helps to democratize astronomical data and knowledge by allowing equal data access to all users. For example, only a few dozen content experts across academia and science centers have recorded tours curated within WWT or have organized WWT-related outreach efforts, yet WWT has reached millions\footnote{Over 10 million unique Windows Application downloads from 2007-2014}: from explorers using the WWT application; to students visiting planetariums and museums around the world; to students at all levels using WWT to augment their curriculum. WWT may be the only visualization software that when astronomers add data and context, it becomes available to students--who may be geographically distant from a science center--to also investigate.

In what follows, we overview the technical aspects of WWT in Section \ref{sec:wwt-overview}, describe some major uses of this visualization tool in Section \ref{sec:wwt-inuse}, outline the open source management and sustainability plan in Section \ref{sustainability}, and conclude in Section \ref{sec:conc}.

\section{The WorldWide Telescope Ecosystem Overview}
\label{sec:wwt-overview}
In the nearly 10 years of WWT development, several applications, services, and technologies were created in order to deliver astronomical data from around the globe in a seamless visualization and narrative environment. In this section, we outline the main components of the WorldWide Telescope software ecosystem.

\subsection{Clients}
WWT is offered as a stand alone Microsoft Windows application and Web-based applications and services. All of the code base is open-sourced under the MIT license and hosted on GitHub\footnote{See the WWT meta-repository: \url{https://github.com/WorldWideTelescope/wwt-home}}.

\subsubsection{Windows Application}
The Microsoft Windows application contains the richest set of features and customizations. It currently serves as the basis of features that are ported to Web application. The WWT Windows application can run natively or through virtualization on either a local virtual machine or on the Cloud.

Beyond the main tasks of visualizing the sky and 3D universe, the WWT Windows application features flexible display modes, supports a variety of controllers, and can incorporate a wide variety of data.

WWT can be displayed in power-walls (ultra high resolution displays created by connecting several monitors or projectors), 3D-stereo projections, virtual reality headsets, interactive kiosk exhibits, and multi-channel environments (e.g., in multiple projector full-dome planetariums). Due to the high demand on computational resources, there are limited plans to port these flexible display modes to the Web application in the near future.

Users may drive WWT with their keyboard and mouse or customize a MIDI controller, an XBOX controller, or the Microsoft Kinect.

A wide variety of data layers are both customizable and controllable. Users can import 3D Models, tabular data, orbital data, imagery (including all-sky maps and planet surface maps), planetary digital elevation models, and catalogs. We discuss the WWT data files in more detail in Section \ref{sec:data-files}.

\subsubsection{Web Application}
A growing subset of features from the WWT Windows application are now delivered via a plug-in free browser-based platform rendered by WebGL. The web application uses the same curated image and catalog data as the Windows application (see Section \ref{sec:data-files}), it supports built-in tour authoring and playback, including a simple slide based or complex key-framed animations. The web application can display tiled images, tours with audio and imagery, and most of the WWT intrinsic data layers.

The WWT Web application leverages the Web Control API for rendering. To better serve astronomical researchers, current development efforts lead by the American Astronomical Society are focused on porting more data layer features of the Windows application to the web application and exposing them in the Web Control API (see Section \ref{sec:apis}).

\subsubsection{Extensibility of WWT Clients}
\label{sec:extensibility}
Users can extend and customize their WWT client experience at a variety of experience levels. They can create a slide-based tour with linked slides (i.e, similar to a DVD menu). With the Windows application users can have full control over WWT tour rendering by using the time-line function (see \href{https://worldwidetelescope.gitbooks.io/worldwide-telescope-advanced-guides/content/timelineeditor.html}{this guide}). They can add their own data layers (see Table \ref{table:data-files} for a list of supported formats). And they can configure a planetarium with multiple computers and projectors to display WWT (see \href{https://worldwidetelescope.gitbooks.io/worldwide-telescope-multi-channel-dome-setup/content/}{this guide}). Below we mention two more means to be extend the WWT clients, by using ``communities'' and by contributing documentation.

\paragraph{Communities} are user-uploaded WWT content (e.g., tours and data layers) that are available to all users of the WWT clients though the Communities tab. Communities content can also be viewed on the WWT website without launching the web application (see \href{https://worldwidetelescope.gitbooks.io/worldwide-telescope-advanced-guides/content/usingcommunities.html}{this guide}).

\paragraph{Documentation} All WWT documentation is now user contributed and maintained on GitHub\footnote{See the WWT documentation meta-repository \url{https://github.com/WorldWideTelescope/wwt-documentation}.}. Each guide is rendered from Markdown as a GitBook\footnote{http://gitbook.com/}, which is available to read online or download in PDF and eBook formats. There are also a growing number of user-contributed WWT tutorials on YouTube. WWT and the WWT Ambassadors (see Section \ref{sec:formal-ed}) maintain YouTube channels (\href{https://www.youtube.com/channel/UCjx-BjxGX6vvxALi9ziBfjA/featured}{WWT Videos}, \href{https://www.youtube.com/user/WWTAmbassadors}{WWT Ambassadors Videos}). We further discuss contributing documentation in Section \ref{sustainability}.

\subsection{Web Services}

\paragraph{Embeddable Web Control\footnote{available at http://worldwidetelescope.org/Use/Embed}} crosses the line between a flexible web application and a software development kit (SDK). It implements a (growing) subset of the WWT Web user interface to support visualizing data sets or playing tours. It is designed to be cut-and-pasted into websites outside of the worldwidetelescope.org domain.

The embeddable code-generation tool creates a WWT namespace on the containing page and spawns an {\tt iframe} which hosts the control. The {\tt webcontrol.aspx} script reads from the options in the containing {\tt div} and passes them to the {\tt iframe} via a {\tt postMessage} API (for cross-domain scripting).

\paragraph{Show Image Service} enables an image to be shared by embedding information about the image in a URL, see the \href{https://worldwidetelescope.gitbooks.io/worldwide-telescope-data-files-reference/content/sharingviews.html#show-image}{documentation} for more details and Table \ref{table:showimage} for examples of its usage.

\subsection{APIs}
\label{sec:apis}
WWT has several application programming interfaces (APIs) of interest to Web and Windows developers. While all of the WWT code base is open sourced, it is incredibly complex, and above the software engineering level of most astronomers and astronomical software engineers. We expect most developers interested in extending or customizing WWT to do so through the APIs (see Section \ref{sustainability} for more discussion on code contributions).

\subsubsection{Windows APIs}
\paragraph{The Layer Control API (LCAPI)} \label{lcapi} allows local or remote applications to send data and control data layers in the Windows application for integration in scenarios outside of the worldwidetelescope.org domain. There is deep integration with the LCAPI and Microsoft Excel through an Excel Add-In tied to the LCAPI, as well as python bindings (see Section \ref{sec:pywwt}).

\paragraph{Windows Client Socket API} is open-sourced, and used by WWT Kinect control to interact with WWT.

\subsubsection{Web APIs}
\label{webcontrol-api}
\paragraph{The Web Control API} is a complex and robust engine that renders WWT objects and imagery into a canvas element. Its purpose is to deeply integrate WWT visualization and tour playback into websites outside of worldwidetelescope.org. It was written in C\# and compiled into javascript using scriptsharp (See the \href{https://worldwidetelescope.gitbooks.io/worldwide-telescope-web-control-script-reference/content/}{API Documentation}). Table \ref{table:webcontrol-api} lists examples of Web Control API usage.

\begin{deluxetable*}{p{1.in}p{4in}p{1in}}
\tablecolumns{3}
\tablecaption{Supported Data Formats\label{table:data-files}}
\tablehead{
\colhead{Name} &
\colhead{Description} &
\colhead{Reference or Documentation}}
\startdata
WTML (also called a ``Collection File'') & XML-coded metadata description of how to find places and streaming data services. & \href{https://worldwidetelescope.gitbooks.io/worldwide-telescope-data-files-reference/content/collections.html}{Collections Guide} \\
\multirow{2}{*}{Tiled Data} & Tiled (quad tree) multi-resolution images sets in a variety of projections delivered through HTTP streams and described by metadata in a WTML file. & \href{https://worldwidetelescope.gitbooks.io/worldwide-telescope-projection-reference/content/}{Projection Reference} \\
& {\it Supported Projections}
\begin{enumerate}
\item  TOAST: Tessellated Octahedral Adaptive Subdivision Transform (TOAST; McGlynn et al., in prep) is a Hierarchical Triangular Mesh \citep[HTM;][]{Kunszt2000} based spherical projection without singularities at the poles and delivered through a quad-tree tile structure.
\item Mercator: Mercator-projected tiles (such as those provided by Bing and Google maps), for all-sky data, the poles are cut off.
\item Equirectangular (plate carr\'ee): equidistant cylindrical projection for all-sky images, the poles have a singularity.
\item Tangential Study Maps: TAN-projected sections of small areas of the sky.
\end{enumerate} & \\
VAMP & VAMP encoded images can be loaded and viewed directly, or tiled by our web service. & \href{https://www.virtualastronomy.org/project.php}{Virtual Astronomy Multimedia Project Standards} \\
3D models & {\tt .3ds} or {\tt .obj} 3D file formats including textures and lighting. & \href{https://worldwidetelescope.gitbooks.io/worldwide-telescope-advanced-guides/content/3dmodels.html}{Guide for Adding 3D Models} \\
Tabular Data & Either through cut/paste or import, tabular data can be visualized through an interactive user interface on the Windows application.  & ... \\
WMS & Web Mapping Service source for tiled and time-series data. & \href{https://worldwidetelescope.gitbooks.io/worldwide-telescope-advanced-guides/content/addingwmsdata.html}{Guide for adding WMS Data}\\
VO Table & Tabular Data in VO Table Format can be viewed as tables or visualized  & \href{http://www.ivoa.net/documents/VOTable/}{VOTable Format Definition} \\
FITS & Flexible Image Transport System (FITS) files in TAN-projection with World Coordinate System (WCS) in celestial coordinates can be viewed and stretched  & \href{https://fits.gsfc.nasa.gov/fits_standard.html}{FITS and WCS Standard}\\
Shapefiles and WKT & Shape files can be loaded and viewed and converted to Well Known Text (WKT) \& Tables. WKT can be included in tabled for viewing complex geometry along with a text extension for displaying oriented text in 3D  & \href{http://www.esri.com/library/whitepapers/pdfs/shapefile.pdf}{ESRI Shapefile Technical Description}\\
ODATA & Tabular feeds from Open Data (ODATA) sources can be mapped in layers and dynamically refreshed at load time.  & \href{http://docs.oasis-open.org/odata/odata/v4.0/odata-v4.0-part1-protocol.html}{Open Data Specification}\\
TLE & Two Line Element orbital data can be used for bulk display of orbits, or as a foundation for a reference frame.  & \href{https://worldwidetelescope.gitbooks.io/worldwide-telescope-advanced-guides/content/minororbits.html}{Guide for adding TLE Data} \href{https://spaceflight.nasa.gov/realdata/sightings/SSapplications/Post/JavaSSOP/SSOP_Help/tle_def.html}{Definition of TLE} \\
SPICE \citep{1996PSS...44...65A,acton2017look} & NASA SPICE kernel data can be imported to display spacecraft or planetary emphemerides by minor manipulation within the NASA SPICE GUI and Microsoft Excel into an {\tt .xyz} text file. & \href{https://astrodavid.gitbooks.io/importing-spice-kernel-data-to-worldwide-telescop/content/}{Guide for adding NASA SPICE Data} \href{https://naif.jpl.nasa.gov/naif/aboutspice.html}{About SPICE}\\
\enddata
\end{deluxetable*}

\begin{deluxetable*}{p{2in}p{4in}}
  \tablecolumns{2}
  \tablecaption{Show Image Service Usage Examples\label{table:showimage}}
  \tablehead{
  \colhead{Name, Website} &
  \colhead{Instructions}
  }
  \startdata
\href{http://montage.ipac.caltech.edu/docs/WWT/}{Montage} & Instructions on Using Montage to TOAST Images for WWT \\
\href{http://chandra.harvard.edu/photo/}{Chandra Photo Album} & Click on an image (from 2012 or earlier), click on the right panel ``View on the Sky (WWT)''  \\
\href{http://www.spacetelescope.org/}{Hubble Space Telescope} & Click on an image and scroll to the bottom right ``View in WorldWide Telescope.'' \\
\href{http://nova.astrometry.net/}{Astrometry.net} & Click on or solve an image and scroll to the bottom right ``View in WorldWide Telescope.'' (Also available as a Flickr plug-in.) \\
\href{http://astropix.ipac.caltech.edu/}{Spitzer IPAC AstroPix} & Click on an image and click ``View in WorldWide Telescope'' on the top right. \\
\enddata
\end{deluxetable*}

\begin{deluxetable*}{p{1.5in}p{3.5in}p{1.in}}
  \tablecolumns{2}
  \tablecaption{Web Control API Examples\label{table:webcontrol-api}}
  \tablehead{
  \colhead{Name, Website} &
  \colhead{Description} &
  \colhead{Supplementary Code}
  }
  \startdata
  \href{http://www.worldwidetelescope.org/GetInvolved/ImportImage}{Import Image} & Import an AVM (Astronomy Visualization Metadata Standard) tagged image or solve for AVM tags with \href{nova.astrometry.net}{astrometery.net} and display it in the WWT Web Application  & \href{https://github.com/WorldWideTelescope/wwt-website/blob/master/WWTMVC5/Scripts/pages/ImportImage.js}{ImportImage.js} \\
  \href{http://www.worldwidetelescope.org/GetInvolved/GreatObservatories}{Great Observatories} & NASA Great Observatories overview and image interactive  &  \href{https://github.com/WorldWideTelescope/wwt-website/blob/master/WWTMVC5/Scripts/pages/Observatories.js}{Observatories.js} \\
  \href{http://www.worldwidetelescope.org/GetInvolved/PlanetExplorer}{Planet Explorer} & Travel from planet to planet in the 3D Solar System mode  &  \href{https://github.com/WorldWideTelescope/wwt-website/blob/master/WWTMVC5/Scripts/pages/PlanetExplorer.js}{PlanetExplorer.js} \\
  \href{http://www.worldwidetelescope.org/GetInvolved/Spectrum}{Spectroscopy} & Spectroscopy overview and interactive   &  \href{https://github.com/WorldWideTelescope/wwt-website/blob/master/WWTMVC5/Scripts/pages/Spectroscopy.js}{Spectroscopy.js} \\
  \multicolumn{3}{c}{Example Web Control API Usage Around the Web} \\
  \href{http://www.spitzer.caltech.edu/glimpse360/wwt}{GLIMPSE 360 WWT Viewer} & ``An infrared view of the disk, or plane, of the Milky Way galaxy assembled from more than 2 million snapshots taken over the past 10 years by NASA's Spitzer Space Telescope.'' &  ... \\
  \href{http://planck.ipac.caltech.edu/wwt/}{Interactive Planck Viewer} & ``...Compare the various components of the sky derived from Planck observations, as well as some other all-sky datasets.''  & ... \\
  \href{http://www.adsass.org/wwt/}{ADS All Sky Survey} & ``The ADS All-Sky Survey is an ongoing effort aimed at turning the NASA Astrophysics Data System (ADS), widely known for its unrivaled value as a literature resource for astronomers, into a data resource.''  &
  \href{https://github.com/adsass/wwt-frontend}{ADSASS WWT-frontend} \\
  \href{http://milkyway3d.org}{Milky Way 3D} & ``A tool intended to organize and curate links to information about data sets relevant to our 3D understanding of the Milky Way'' & ... \\
  \href{http://cxc.harvard.edu/csc2/wwt.html}{``'Chandra Source Catalog Release 2.0} & The Chandra Source Catalog (CSC) is ultimately intended to be the definitive catalog of X-ray sources detected by the Chandra X-ray Observatory. To achieve that goal, the catalog will be released to the user community in a series of increments with increasing capability. The first official release of the CSC 2.0 will include 10,382 Chandra ACIS and HRC-I imaging observations released publicly through the end of 2014. When complete, CSC 2.0 will provide information for about 370,000 unique sources in several energy bands.'' & ... \\
  \href{https://pywwt.readthedocs.io/en/latest/}{pyWWT} & The pyWWT package aims to make it easy to use WorldWide Telescope from Python (see Section \ref{sec:pywwt}).
& \href{https://github.com/WorldWideTelescope/pywwt}{pyWWT on GitHub}
  \enddata
\end{deluxetable*}

\paragraph{Web Services API} – Sets of hosted web services that deliver or process data to and for the WWT clients and other consumers, such as the ShowImage service referenced in Table \ref{table:showimage}. Related documentation is hosted on \href{https://worldwidetelescope.gitbooks.io/worldwide-telescope-data-files-reference/content/sharingviews.html}{GitHub}.

\subsection{Tools and SDKs}
While part of Microsoft Research, WWT published software development kits (SDKs) to assist users and developers with importing data (available \href{http://www.worldwidetelescope.org/GetInvolved/Develop#sdks}{here}). These tools are in the process of being converted to web services to support all operating systems.
\paragraph{Sphere Toaster} takes small to medium sized images and outputs Tessellated Octahedral Adaptive Subdivision Transform (TOAST; McGlynn et al., in prep) tile pyramids, plate files, and WTML.
\paragraph{Study Chopper} takes small to medium study images and outputs TAN-projected tile pyramids and WTML.
\paragraph{AVM Import Tool} takes Astronomy Visualization Metadata Standard-tagged images with WCS coordinates and creates hosted tile pyramids and WTML files to access them. This service is available online \href{http://www.worldwidetelescope.org/GetInvolved/ImportImage}{here}.
\paragraph{Tile SDK} A software tool kit with samples that allows users to create a custom data transformation pipeline to take images or elevation data from input format and output it to WWT compatible pyramids.
\paragraph{Terapixel Project} While not exactly an SDK, the Terapixel Project is an open source pipeline that was used to reprocess the original DSS data from scanned plates into a Terapixel TOAST pyramid with globally optimized color and brightness corrections \citep{Agarwal2011}.

\subsection{Curated Data}
\label{sec:data-files}

The data shown WorldWide Telescope is a mix of curated data (stored on the Azure Cloud) and data displayed from around the globe, including the National Virtual Observatory (NVO) registries. There are currently more than 90 all-sky surveys, 20 planet maps, and 60 panoramas (see Appendix \ref{sec:appendix}: Tables \ref{table:allsky-data}, \ref{table:panorama-data}, and \ref{table:planet-data}). In addition, WorldWide Telescope hosts over 500 contributed images from education and public outreach offices at Hubble, Chandra, Spitzer, Gemini, NOAO, NRAO, as well as from astrophotographers.

\subsection{Supported Data Formats}

\begin{figure*}
  \includegraphics[width=.48\textwidth]{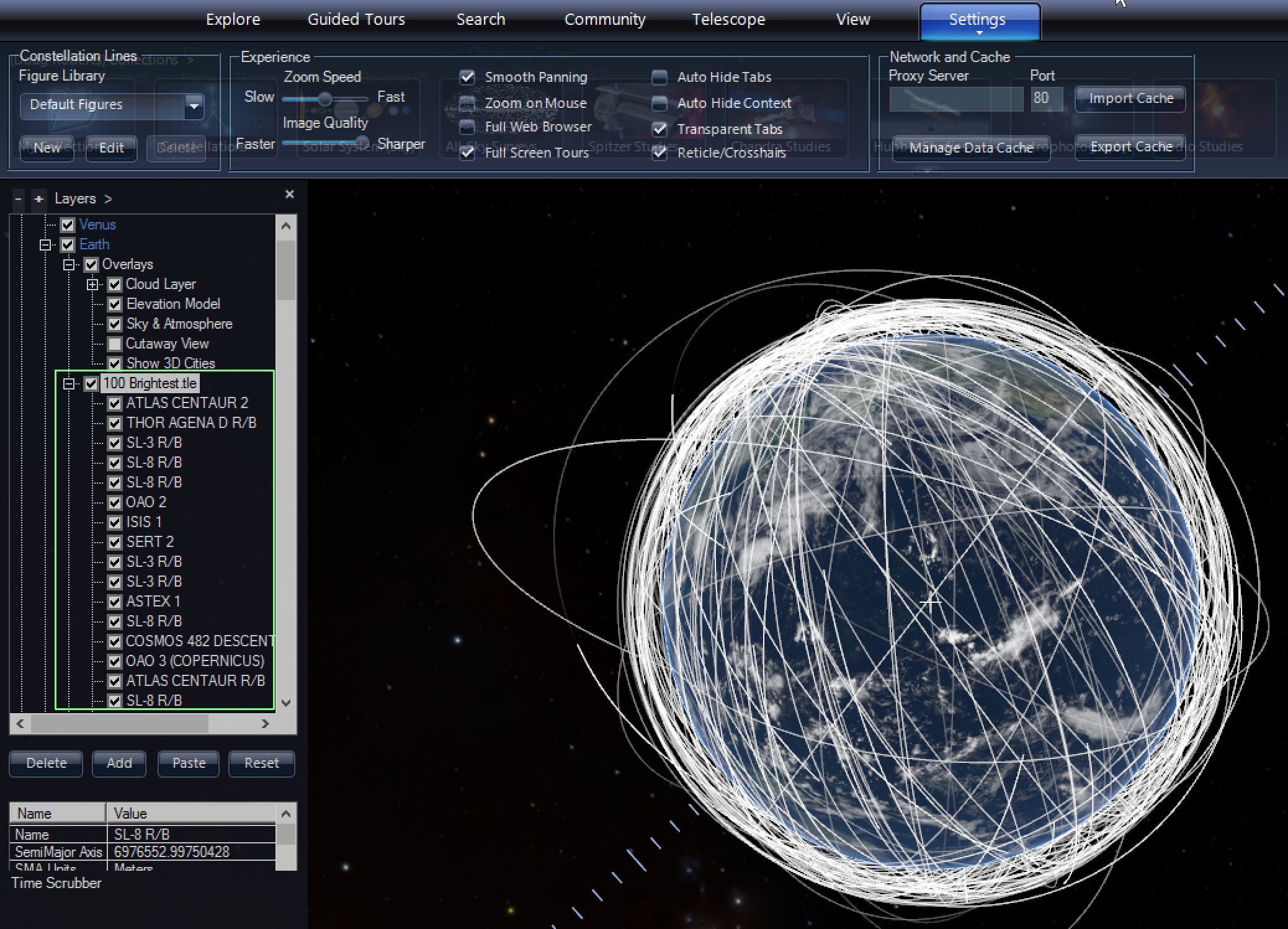}
  \includegraphics[width=.48\textwidth]{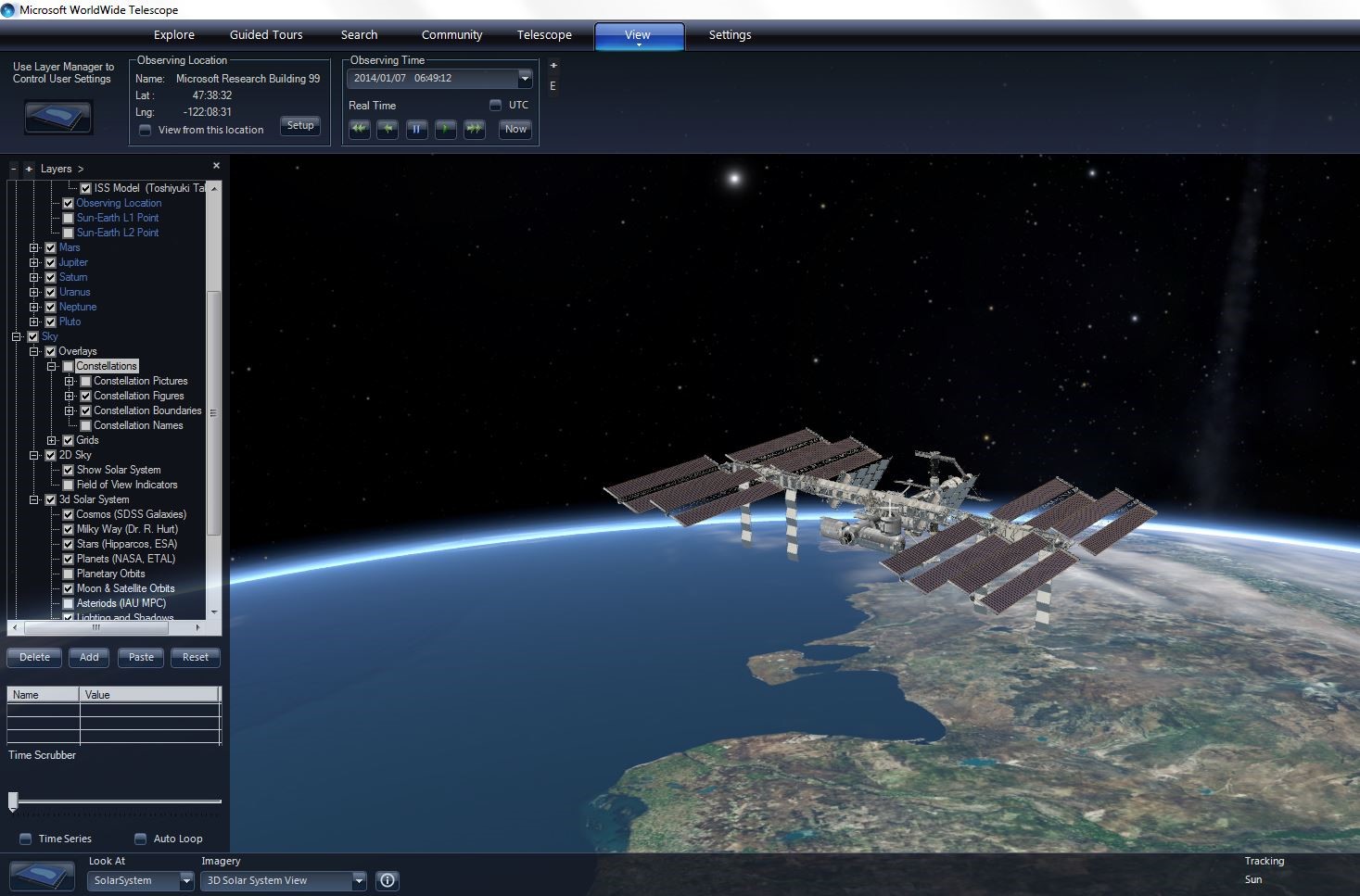} \\
  \includegraphics[width=.48\textwidth]{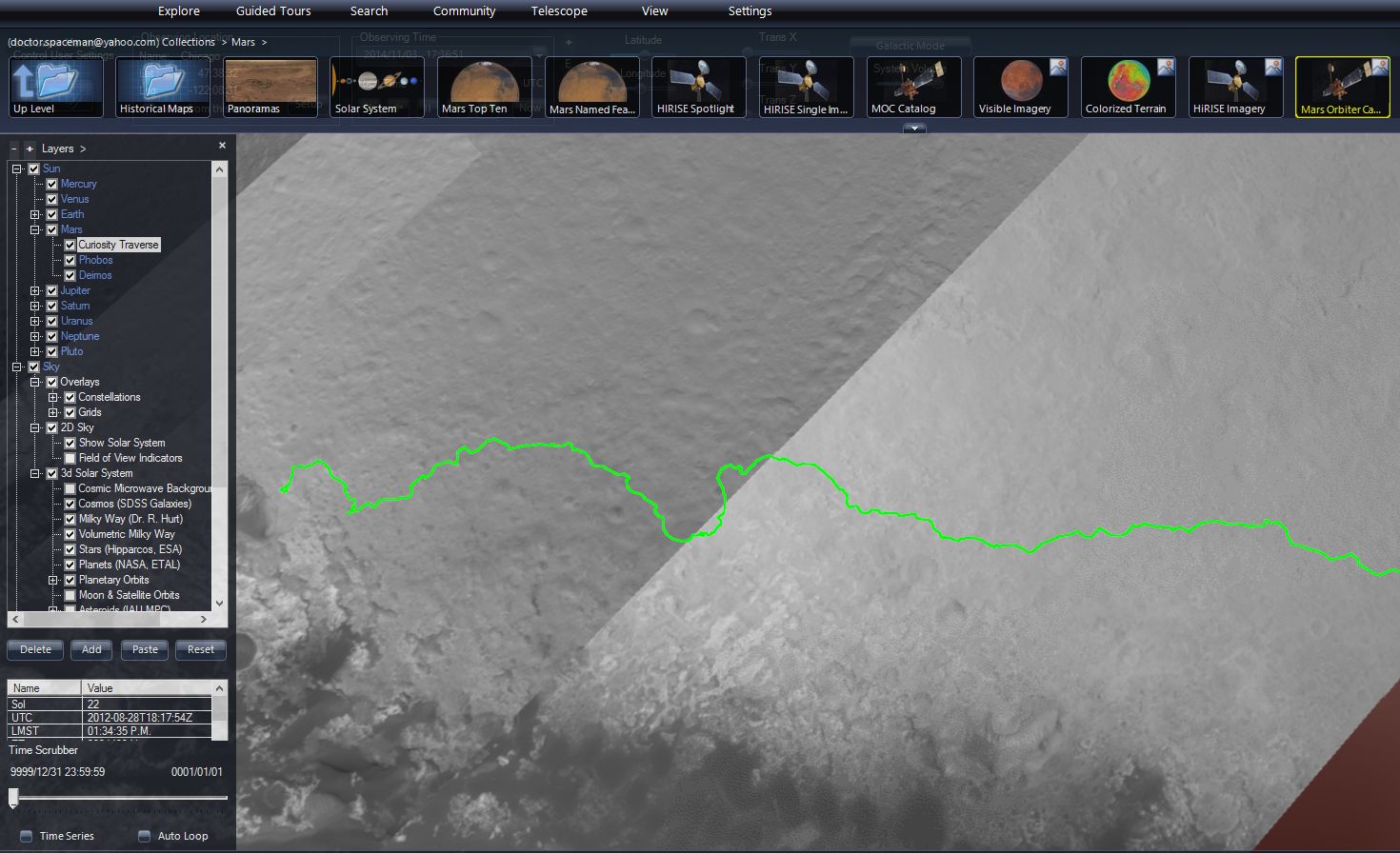}
  \includegraphics[width=.48\textwidth]{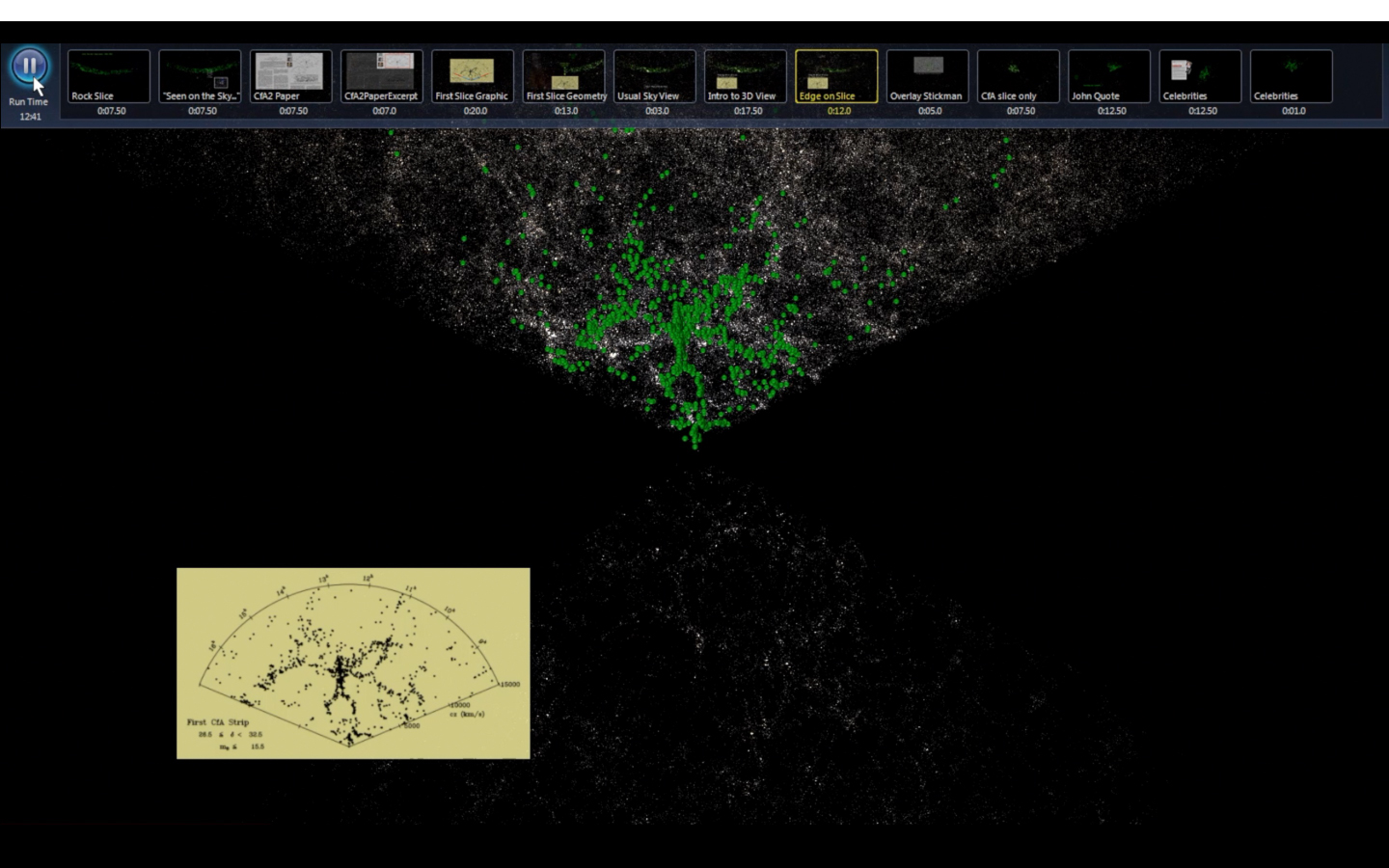}
  \caption{Examples of supported data and models. Clockwise from top left: View of Earth with near Earth objects imported as Two Line Element (TLE) files (data from International Astronomical Union's Minor Planet Center); 3D ISS Model (ISS Model from: Toshiyuki Takahe). A 3D rendering of a user-plotted catalog \citep[The CfA Redshift Survey, green;][]{Geller1989} over the COSMOS dataset \citep{Scoville2007}; Latitude, Longitude, and Altitude of a Mars Rover imported to WWT using the WWT-Excel plug-in (data courtesy of Joe Knapp, \url{curiosityrover.com}). \label{fig:data}}
\end{figure*}

WWT supports a variety of data formats (see Figure \ref{fig:data}), which users can import via the Windows application. Importing of data on the Web application is in development, natively it currently supports FITS images, and many more data formats are possible through an astropy-affiliated package (see Section \ref{sec:pywwt}). Table \ref{table:data-files} lists current levels of data support.

\subsubsection{Plate Files}
The reasoning behind splitting large images into tiles is to read from disk only as much as is absolutely necessary for each database query (e.g., a view of a portion of the sky). The efficiency that tiling enables gives way in the limit of reading and transferring from the server, many small files. The WWT team created ``plate files'' as an efficient way to store, exchange and deliver tiled multi-resolution images. Plate files have an index and data storage and allow millions of small files to be packed into a single file. Single large files copy up to 2 orders of magnitude faster than the millions of files they contain. There exist two plate file formats:
\begin{enumerate}
  \item For densely packed complete surveys or studies with single generation data.
  \item For sparse data that can contain multiple generations of data for a given tile index. This format is great for building large surveys over time with streaming data, for example, the Mars HiRise data.
\end{enumerate}

\section{WorldWide Telescope Usage}
\label{sec:wwt-inuse}
WWT has been used extensively at nearly all levels of astronomy education. For clarity, we separate the following section by formal and informal education, and include education and public outreach (EPO) centers as part of informal education.

\subsection{Formal Education}
\label{sec:formal-ed}
The most prominent formal education program that uses WWT in the WorldWide Telescope Ambassador Program\footnote{https://wwtambassadors.org} (WWTA) which has evolved alongside WWT since 2010.

The WorldWide Telescope Ambassadors Program aims to educate the public about Astronomy and Science using WWT.  It is run by a team of astronomers and educators at Harvard University, in collaboration with the AAS and Microsoft Research. They recruit and train volunteer ambassadors who help facilitate the use of WWT in educational settings like schools, science festivals, and museums.

\subsubsection{Pre-College Courses}
WWTA have directly served thousands of students in the classroom.  Since 2015, more than 700 of these students had extensive learning experiences that included one or two full weeks of instruction from WWTA team members \citep[about Moon phases, seasons, and life in the Universe, see][]{Udomprasert2014,Udomprasert2016}. They have pre-post assessment data showing outstanding learning gains from these interventions \citep[][Udomprasert et al. in prep]{Udomprasert2012}. Lesson plans and other materials are available on their \href{https://wwtambassadors.org/science-education-research}{website}.

WWT has also been used for Astronomy classes at Beijing Shijia Primary School for three years. A Guidebook on Interactive Astronomy Teaching (Primary School Teacher version) designed by Chinese Virtual Observatory project is published by Popular Science Press in Beijing.

\subsubsection{College Courses and Beyond}
\paragraph{Astronomy Courses} The WWTA program has helped to create WWT Windows Application-based introductory astronomy labs on parallax and Hubble's law. They have been used at Bucknell University and the University of Massachusetts Amherst \citep[see][and the \href{https://wwtambassadors.org/astronomy-101}{WWTA Website}]{Ladd2015}.

WWT has been part of introductory astronomy courses at the Central China Normal University since 2009 \citep{Qiao2010, Wang2015}, and a graduate ISM course at Harvard University \citep{Sanders2014}. Co-author Weigel uses WWT to visualize spacecraft ephemerides and planetary texture maps in context in an immersive fulldome environment for the Introduction to Astronomy class at Samford University. These visuals are presented and manipulated in real time in the Christenberry Planetarium on Samford's campus.

\paragraph{General Science Courses} Samford University makes further use of WWT in their Cultural Perspectives and Scientific Inquiry classes. Weigel teaches about the Scientific Revolution using WWT in a planetarium. Students are guided on a journey from naked eye astronomical observations and antiquated methods of scientific reasoning and philosophy, through the age of the scientific revolution and the subsequent invention of telescopes, and finish with a look at current big data astronomy and predictions for the future.

\paragraph{Science Communication} The University of Washington offered an undergraduate seminar on science communication as applied to creating a tour in WWT (curriculum is available \href{http://philrosenfield.github.io/teaching_files/tourmaking/CourseGuide.pdf}{here}). Students learned story boarding, distilling vs. dumbing down, avoiding jargon, and how to create WWT tours playable in the University of Washington planetarium.

\subsection{Informal Education}

WWT has shown to be an excellent data-driven astronomical education and public outreach environment. For example, it inspired co-author Cui to initiate the IAU Working Group ``Data driven Astronomical Education and Public Outreach'' \citep[DAEPO;][]{Cui2018}, it has been used to help capture and share indigenous Australian astronomical knowledge\citep{Nakata2014}, and has allowed planetariums to be `flipped,' allowing students to present planetarium shows to their peers.

\subsubsection{Museums and Planetariums}
\begin{figure}
  \includegraphics[width=\columnwidth]{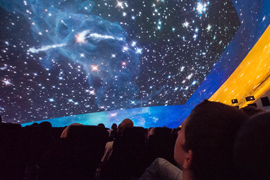}
  \caption{Adler Planetarium Audience viewing WWT during Cosmic wonder credit: Microsoft Research Blog \label{fig:cosmicwonder}}
\end{figure}

WWT has been used for live and recorded shows in major planetariums, such as the Adler Planetarium (\href{http://www.adlerplanetarium.org/events/cosmic-wonder/}{Cosmic Wonder}; Figure \ref{fig:cosmicwonder}) and the Morrison Planetarium (\href{https://www.calacademy.org/press/releases/earthquake-planetarium-show-opens-on-may-26-2012-at-the-california-academy-of}{Earthquake}).

WWT has also proven to be an attractive choice for smaller planetariums, since the software is free and it comes with a robust projector calibration system\footnote{See the \href{https://www.gitbook.com/book/worldwidetelescope/worldwide-telescope-multi-channel-dome-setup/details}{Multi-Channel Dome Guide}}. The following list of WWT-driven smaller planetariums is incomplete, but includes the University of Washington, Mount Hood Community College, Central Washington University, Samford University, and Bellevue College. In China, 6 WWT-driven planetariums have been built and 2 are currently under construction, with an Internet-based WWT-driven planetarium alliance for resource and experience exchange.

\begin{figure}
\includegraphics[width=\columnwidth]{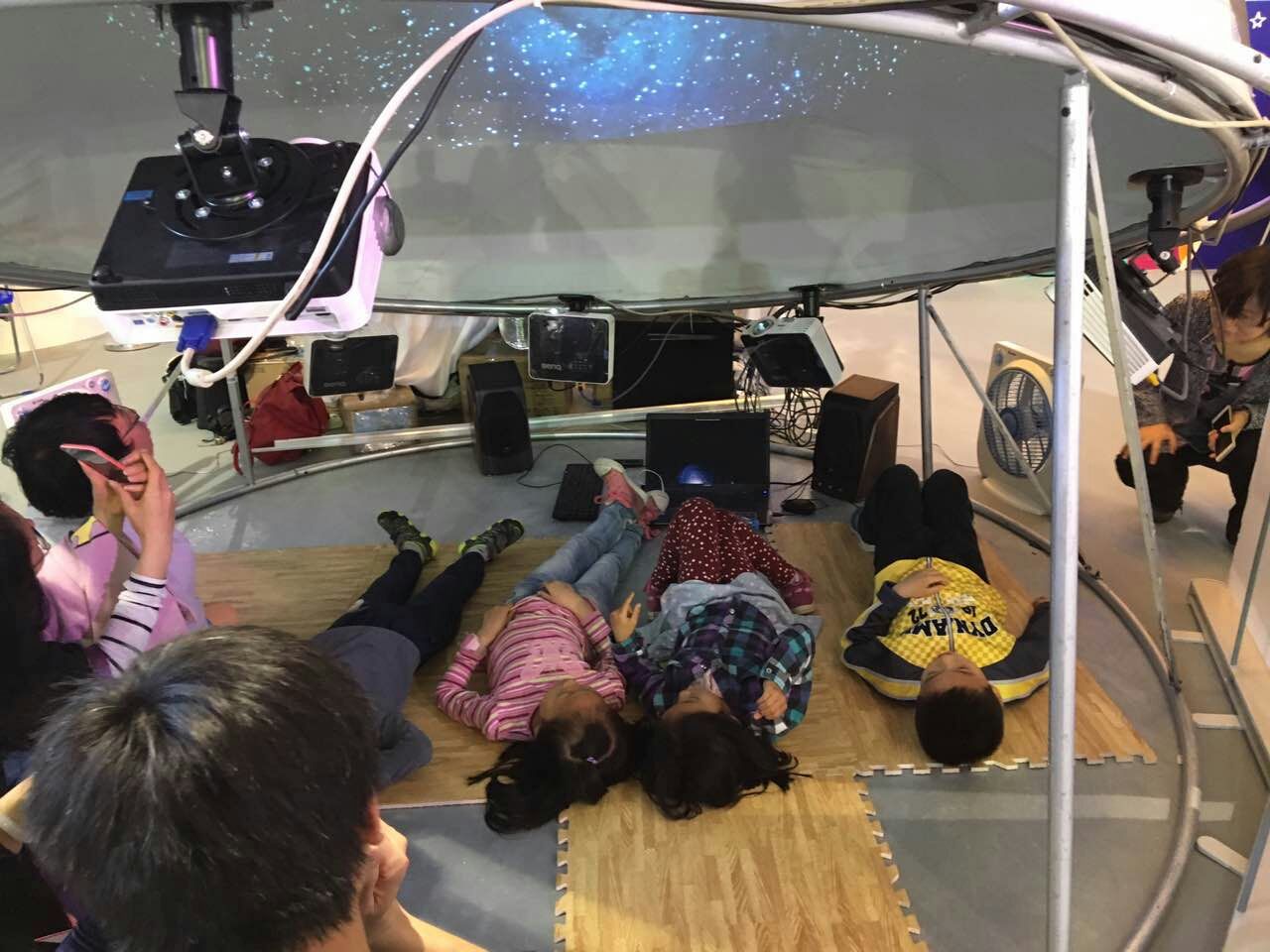}
 \caption{2016 Science and Technology Week (China). Children are watching WWT tour shown inside a portable planetarium. \label{fig:scifest}}
\end{figure}

Mobile, portable, or traveling planetarium programs have also begun to use WWT. For example, Discovery Dome\footnote{\url{http://eplanetarium.com/software_worldwide_telescope.php}} packages WWT software with their inflatable domes, and the University of Washington, the University of Oklahoma (as the Soonertarium), and Harvard University all have WWT-driven mobile planetariums programs. Nearly every mobile planetarium program has participated in some form of a local science festival, for example, Figure \ref{fig:scifest} shows students watching a tour in a portable WWT planetarium.

\begin{figure}
\includegraphics[width=\columnwidth]{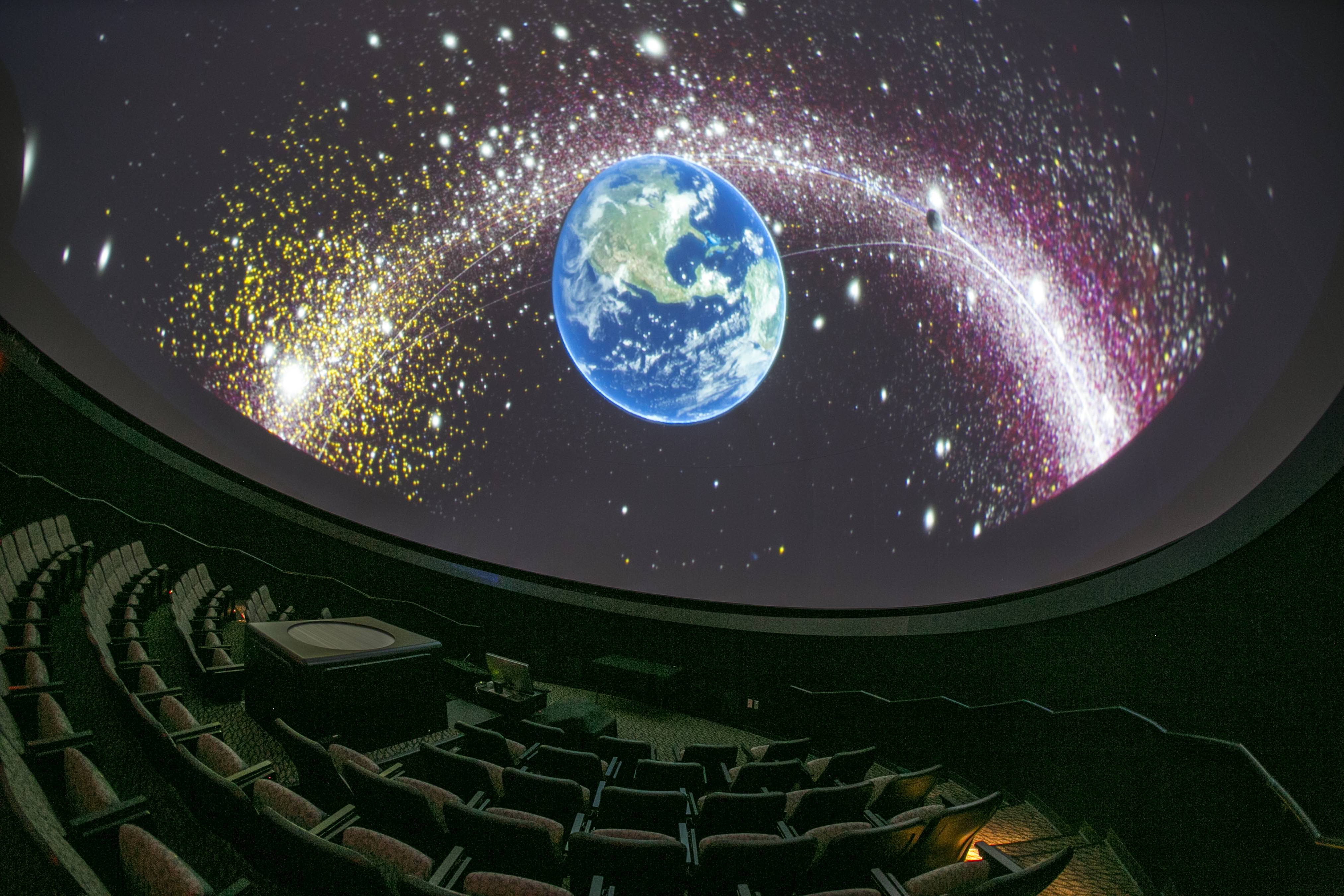}
\caption{WorldWide Telescope rendered in the full dome Christenberry Planetarium at Samford University. Credit: Samford University \label{fig:iur}}
\end{figure}

One of the unique uses of WWT in an educational setting comes from `flipping' the classroom or in this case, the planetarium. Instead of students passively attending a lecture or recorded video in a planetarium, they can research and then create their own content to share in a planetarium. This application of WWT has been independently described by several groups. The DAEPO has coordinated student authored WWT tour contests\footnote{\url{http://english.nao.cas.cn/ns/ConferenceNWorkshop/201706/t20170614_178121.html}} three times. The Cal Academy helped San Francisco Bay Area students to present WWT in the Morrison Planetarium \citep{Roberts2014}. The University of Washington works with Seattle-area middle and high schools to present WWT in a mobile planetarium \citep{Rosenfield2014}. The Christenberry Planetarium at Samford University mentors middle and high school students in summer programs, teaching them to produce and present WWT tours for public audiences in a planetarium (see Figure \ref{fig:iur}).

\begin{figure}
\includegraphics[width=\columnwidth]{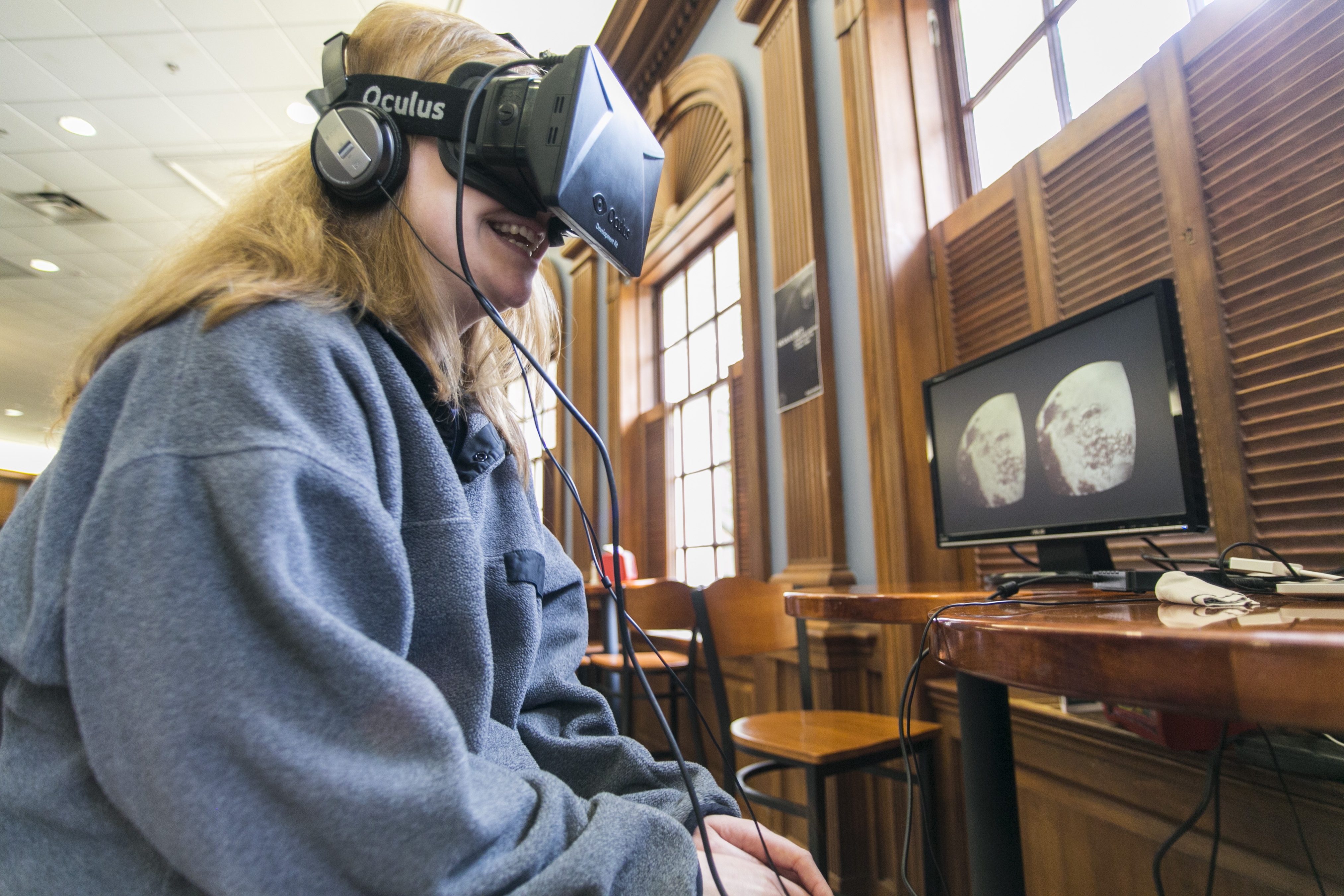}
 \caption{Samford University student experiencing VR over the surface of Pluto (New Horizons) in WWT. Credit: Samford University \label{fig:vrheadset}}
\end{figure}

The Christenberry Planetarium also uses WWT's VR capabilities in public outreach to promote the latest astronomical discoveries and research by adding the ``wow'' factor by using cutting edge technology within the university and the greater Birmingham, Alabama community. Weigel and his students develop VR specific WWT tours and host programs to both inspire and teach youth in astronomy in Alabama (see Figure \ref{fig:vrheadset}).

\begin{figure}
\includegraphics[width=\columnwidth]{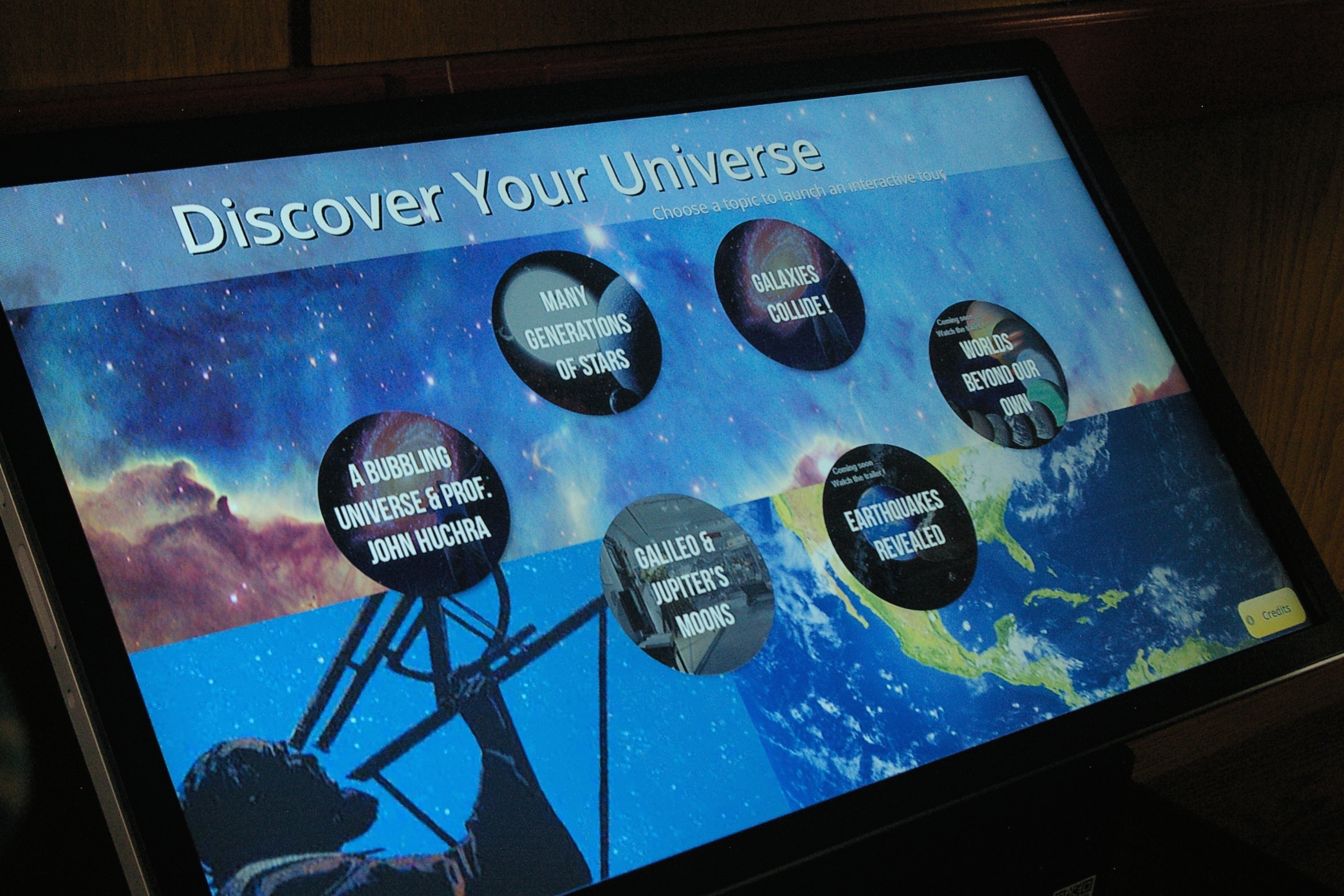}
 \caption{Harvard University's WWT-driven Kiosk \label{fig:kiosk}}
\end{figure}

WWT has a kiosk mode where a user can interact with a guided tour or contained aspect of WWT. An incomplete list of museums and informational centers that have used WWT in this way include Harvard University (see Figure \ref{fig:kiosk}), the Adler Planetarium (see the \href{http://www.adlerplanetarium.org/whats-here/dont-miss/space-visualization-lab/}{Space Visualization Lab}), and the Imiloa Astronomy Center of Hawai'i which had a Microsoft Kinect driven WWT exhibit  \href{https://subarutelescope.org/Gallery/movie_subaru.html}{``The Universe at your Finger tips''}.

\subsubsection{Education and Outreach Centers}
Both WWT clients have served as a conduit for publishing EPO imagery and stories. For example, the ``Explore'' tab shows curated feeds from the Hubble Space Telescope, Spitzer Science Center, Chandra X-ray Observatory, and the European Space Agency. Several outreach efforts have also made use of the WWT Web Control API (see bottom of Table \ref{table:webcontrol-api}) and the WWT Show Image service (see Table \ref{table:showimage}).

\subsection{Astronomical Research}
\label{sec:research}

WWT enters the astronomical research work-flow at several points: discovery, helping visualize big and wide datasets; analysis, WWT APIs allow analysis tools to link data; and dissemination, WWT can enhance figures and presentations, and researchers can create video abstracts to accompany their submitted manuscripts to journals (see Section \ref{sec:dissemination}). WWT and its APIs allow observatories and mission centers to adopt a common visualization interface for their databases. In the following section, we discuss the most recent efforts and future outlook led by the AAS to help the astronomical research community benefit from WWT technologies.

\subsection{Discovery \& Analysis}
\label{sec:pywwt}
Astronomers interested in using survey data (e.g., SDSS, LSST, Gaia, DESS) have most likely shifted from the traditional \textit{ftp-grep} data analysis model, where a scientist first downloads a copy of data (ftp) and then uses an analysis package to identify for patterns, or answer their research question (grep) \citep[see e.g.,][]{Gray2004}. Instead, astronomers analyze their data by querying remote databases, and soon, accessing hosted computing facilities to perform their analysis \citep[e.g, LSST:][and refs. therein]{Juric2015}. WorldWide Telescope was first designed to exist in the post ftp-grep regime, and now that is has been open sourced and ported to the Web, it has plug and play potential for use in smaller-budgeted observatories and data archives.

\paragraph{Archival Centers} The next release of the WWT Web Control API will have data import/export, and selection features which will make it powerful link between data archives and researchers. Observatories and data centers can insert the web control on their website and have a robust, open source, data archive visualization engine. Observatories or data archive centers that use WWT in this way will automatically  benefit their EPO efforts, since EPO staff would have the same access to data and be able to create tours or build a custom user interface over the WWT Web Control API.

In addition, abstract services, such as ADS\footnote{\url{http://adsabs.harvard.edu}} can use the WWT interface as another means for users to discover research in astronomy. The Zooniverse's Astronomy Rewind project\footnote{\url{https://www.zooniverse.org/projects/zooniverse/astronomy-rewind}} is a citizen scientist project to place AAS journal figures dating back to the 1800s into the Astronomy Image Explorer\footnote{\url{http://www.astroexplorer.org}} with AVM tags so they can be viewed in WWT.

\paragraph{Data Analysis Tools}
Several data analysis tools are being built off of the Web Control API. For example, a link between JS9\footnote{\url{https://js9.si.edu}; the online javascript version of DS9 (\url{http://ds9.si.edu)} with a public API} and WWT has been established as part of the AAS Astrolabe project\footnote{The \href{http://astrolabe.arizona.edu/}{Astrolabe Project} is creating a new open-access repository for previously un-curated astronomical datasets, building on existing CyVerse infrastructure with robust cloud-based resources for managing, linking, processing and sharing research data.}. Users of this service can upload their images to JS9, manipulate them as they would in DS9, and view the result in WWT.

Python bindings for the WWT Web Control API are a new focus and were first prototyped in Glue-viz\footnote{Glue-viz is Python library to explore relationships within and among related \url{http://glueviz.org/en/stable/}, see \href{http://glueviz.org/en/stable/whatsnew/0.11.html?\#experimental-worldwide-telescope-plugin}{WWT Plugin}}.

\begin{figure}
\includegraphics[width=\columnwidth]{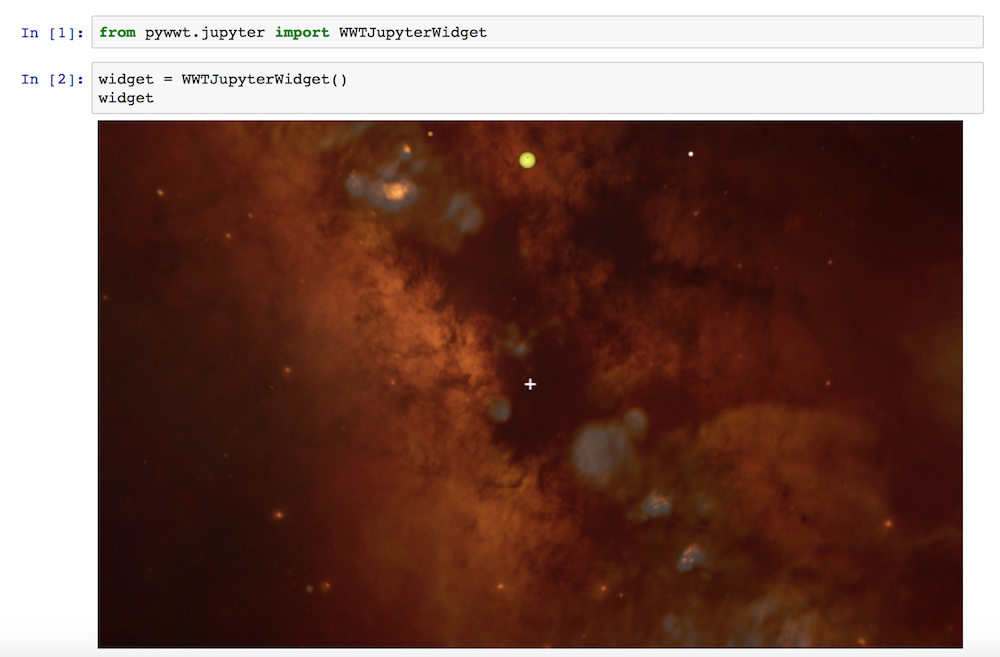}
\caption{A screen shot of pyWWT, a Python interface for WorldWide Telescope based on the Web Control API. \label{fig:jupyter}}
\end{figure}

\paragraph{PyWWT: A Python interface to WorldWide Telescope}
PyWWT is a package that provides access to much of the Windows and web clients' functionality to Python on a variety of platforms. It allows the WWT web client to be used as an interactive widget inside Jupyter Notebooks and and Jupyter Lab \citep[pictured in Figure \ref{fig:jupyter};][]{kluyver2016jupyter}, as well as through a standalone interactive Qt-based viewer/widget\footnote{\url{https://www.qt.io}}, and also provides a Python-based client for the WWT desktop application on Windows (using the LCAPI, see Section \ref{sec:apis}).

In all cases, users can pan and zoom with the mouse, display multiple layers from all-sky surveys, and play tours. PyWWT makes it easy to add annotations, or shapes, that can be used to highlight celestial objects, enhance tours, or illustrate fields of view for telescopes, among many other applications. The package is written to integrate smoothly with Astropy\citep{2013A&A...558A..33A} for easy adaptability to typical research work-flows such as those described in Section \ref{sec:pywwt}. Support for Astropy tables and displaying local FITS images is forthcoming.

PyWWT is under active development and is hosted under the WorldWideTelescope GitHub organization (see Table \ref{table:webcontrol-api}), where its full code base is open-sourced (version 0.3.0 was launched in December 2017).

\subsection{Dissemination}
\label{sec:dissemination}
We have seen examples of the WWT narrative layer (tour features) in sharing stories and teaching astronomy. WWT has also been used in video abstracts, which have converted WWT tours into videos distilling a scholarly publication \citep[see the Publisher's website of ][]{Berriman2017,Batygin2016,Yusef-Zadeh2015,Currie2015}. Future WWT Web Control development will allow for embedding WWT in 2- and 3D modes as an interactive figures in scholarly publications.

\section{Management and Sustainability of WorldWide Telescope}
\label{sustainability}
WorldWide Telescope is managed by the AAS its (MIT) copyright is held by the .NET Foundation. Below we overview the management and sustainability plan for this open source software project.

\subsection{Project Leadership}

AAS Board of Trustees (AASBoT) is in charge of managing the AAS WWT project. The AASBoT has mandated the \href{http://worldwidetelescope.org/About}{AAS WWT Steering Committee} to continually assess the WWT project against metrics based on the WWT's \href{http://worldwidetelescope.org/About}{purpose statement} as well as the AAS' \href{https://aas.org/governance/aas-mission-and-vision-statement}{purpose statement}. The AAS WWT Steering Committee and AAS WWT Project Director work together to interpret the purpose statements into initiatives with quantitative metrics for success.

\subsection{Open Source Contributions}

The longevity and success of open source software depends on the community. We are striving to build a inclusive WWT community. We expect everyone in the WWT community to follow our  \href{https://github.com/WorldWideTelescope/wwt-home/blob/master/conduct.md}{code of conduct} and a \href{https://github.com/WorldWideTelescope/wwt-home/blob/master/CONTRIBUTING.md}{contributing} guide. Below, we discuss two avenues of contribution: contributing to the code base, and contributing documentation and examples.

\subsubsection{Code Contributions}
We maintain a meta-repository, \href{https://github.com/WorldWideTelescope/wwt-home}{wwt-home}, which primarily contains a table of information on the different WWT-related repositories in the WorldWideTelescope GitHub organization. Importantly, each table entry includes a description of the project as well as the skills required to contribute. Until we have a robust user community to review code contributions (as well as proposals for enhancements), they are considered by the \href{http://worldwidetelescope.org/About}{WWT Advisory Board}.

\subsubsection{Documentation Contributions}
As discussed in Section \ref{sec:extensibility}, all of our documentation has been open sourced and hosted on GitHub, accessible via our meta-repository, \href{https://worldwidetelescope.github.io/wwt-documentation/}{wwt-documentation} including \href{https://github.com/WorldWideTelescope/worldwide-telescope-web-control-script-reference/blob/master/samples.md}{code samples}.

Contributers can edit existing documentation or create their own, even if they are not familiar with GitHub or Markdown. In addition, they can create documentation within a GitBook or import files created using Microsoft Word. Each new contribution will be added as a line to the wwt-documentation table which is the basis of the  \href{http://www.worldwidetelescope.org/Learn/UserGuide}{www.worldwidetelescope.org's User Guides Page}.

Sample code can be published as gists and linked to our meta-repositories or as Jupyter notebooks (e.g., \href{https://github.com/WorldWideTelescope/pywwt-notebooks}{pyWWT example notebooks}).

We will continue to engage and consult with the broader astronomical community to plan new features and services via AAS member polls, WWT online forum, tracking GitHub issues, the WWT help desk (wwt@aas.org), and feedback during workshops and webinars (direct and via assessments).

\subsubsection{Contributing WWT Products}
In the future, we hope to create a place on the Web to house WWT-related workshop materials, contributed WWT lesson plans, 3D models, music, and more--that meet the existing skill sets of the contributers.

\section{Conclusions}
\label{sec:conc}
We have reviewed the software ecosystem of WWT and many of its uses in astronomy. WWT is astronomical visualization software what can directly link to all levels of education, in that way, WWT can democratize astronomical data, but only with the help of the astronomical community.

In summary:
\begin{itemize}
  \item WWT is open source (MIT license) and managed by the American Astronomical Society.
  \item WWT has two main applications: 1) a Microsoft Windows application that is a mainstay in planetariums and international education; and 2) a WebGL powered Web application that is operating system agnostic which we expect most US-based astronomers to prefer.
  \item Each application has APIs and services to allow users and developers to customize and extend WWT without needing the high level of C\# programming knowledge to advance the core WWT components.
  \item WWT has been used innovatively in formal and informal educational settings, supplementing: a) early learners physical understanding of the seasons and moon phases; b) college students' introductory astronomy and general science courses as well as at least one graduate astronomy course; f) planetarium education, e.g., by enabling students to present planetarium shows to their peers.
  \item Educational programs have used WWT in a variety of settings: computer labs, kiosks, virtual reality headsets, and planetariums.
  \item Ongoing development of the WWT Web Control API will help to embed the powerful tool in astronomical research: a) as a visualization engine for data archives; b) linked to python and javascript analysis tools; c) interactive (or video) figures for scholarly publications and presentations.
\end{itemize}

\acknowledgements

Authors thank WorldWide Telescope Advisory Board whose effort of to support the WWT project since its inception and help transition WWT from Microsoft to the AAS insured its existence: Sarah Block, Andy Connelly, \href{https://orcid.org/0000-0002-7456-1826}{Chen-Zhou Cui}, Cristine Donnelly, Karl Fogel, \href{https://orcid.org/0000-0001-8246-5001}{Ron Gilchrist}, \href{https://orcid.org/0000-0003-1312-0477}{Alyssa Goodman}, Morgan Griffith, \href{https://orcid.org/0000-0002-4601-8180}{Bryan Heidorn}, \href{https://orcid.org/0000-0002-4347-6901}{Robert Hurt}, \href{https://orcid.org/0000-0003-2500-8984}{Jonathan Fay}, Erin Johnson, \href{https://orcid.org/0000-0001-7482-2814}{Susanna Kohler}, \href{https://orcid.org/0000-0002-7134-8296}{Knut Olsen}, \href{https://orcid.org/0000-0002-7132-418X}{Fred Rasio}, Doug Roberts, \href{https://orcid.org/0000-0001-9306-6049}{Philip Rosenfield}, \href{https://orcid.org/0000-0001-8814-863X}{Gretchen Stahlman}, Mark SubbaRao, \href{https://orcid.org/0000-0002-8596-6634}{Julie Steffen}, \href{https://orcid.org/0000-0002-5294-0198}{Matt Turk}, Patricia Udomprasert, Jaap Vreeling, \href{https://orcid.org/0000-0002-8026-2291}{A. David Weigel}, Ryan Wyatt, Martin Woodward, and Curtis Wong. We also thank the the current and past AAS WWT Steering Committee members for support, ideas, and feedback on development initiatives: Amanda Bauer, \href{https://orcid.org/0000-0002-8704-4473}{Nancy Brickhouse}, \href{https://orcid.org/0000-0002-4468-2117}{Jack Burns}, Buell Jannuzi, Jessica Kirkpatrick, \href{https://orcid.org/0000-0003-3217-5967}{Sarah Loebman}, \href{https://orcid.org/0000-0002-3972-8790}{Kevin Marvel}, \href{https://orcid.org/0000-0003-4365-1455}{Megan Schwamb}, and \href{https://orcid.org/0000-0002-5858-6767}{Chick Woodward}.  We thank Doug Roberts and \href{https://orcid.org/0000-0002-8596-6634}{Julie Steffen} for AAS WWT project leadership.

Chen-Zhou Cui thanks support from National Natural Science Foundation of China (NSFC)(11503051, U1531111, U1531115, U1531246, U1731125, U1731243). Chen-Zhou Cui also thanks the National R\&D Infrastructure and Facility Development Program of China, ``Earth System Science Data Sharing Platform'' and ``Fundamental Science Data Sharing Platform'' (DKA2017-12-02-XX).

This publication makes use of data products from the Two Micron All Sky Survey, which is a joint project of the University of Massachusetts and the Infrared Processing and Analysis Center/California Institute of Technology, funded by the National Aeronautics and Space Administration and the National Science Foundation.

We acknowledge the use of the Legacy Archive for Microwave Background Data Analysis (LAMBDA), part of the High Energy Astrophysics Science Archive Center (HEASARC). HEASARC/LAMBDA is a service of the Astrophysics Science Division at the NASA Goddard Space Flight Center.

Funding for the SDSS and SDSS-II has been provided by the Alfred P. Sloan Foundation, the Participating Institutions, the National Science Foundation, the U.S. Department of Energy, the National Aeronautics and Space Administration, the Japanese Monbukagakusho, the Max Planck Society, and the Higher Education Funding Council for England. The SDSS Web Site is \url{http://www.sdss.org/}. The SDSS is managed by the Astrophysical Research Consortium for the Participating Institutions. The Participating Institutions are the American Museum of Natural History, Astrophysical Institute Potsdam, University of Basel, University of Cambridge, Case Western Reserve University, University of Chicago, Drexel University, Fermilab, the Institute for Advanced Study, the Japan Participation Group, Johns Hopkins University, the Joint Institute for Nuclear Astrophysics, the Kavli Institute for Particle Astrophysics and Cosmology, the Korean Scientist Group, the Chinese Academy of Sciences (LAMOST), Los Alamos National Laboratory, the Max-Planck-Institute for Astronomy (MPIA), the Max-Planck-Institute for Astrophysics (MPA), New Mexico State University, Ohio State University, University of Pittsburgh, University of Portsmouth, Princeton University, the United States Naval Observatory, and the University of Washington.

The Digitized Sky Survey was produced at the Space Telescope Science Institute under U.S. Government grant NAG W-2166. The images of these surveys are based on photographic data obtained using the Oschin Schmidt Telescope on Palomar Mountain and the UK Schmidt Telescope. The plates were processed into the present compressed digital form with the permission of these institutions.

This research has made use of the USNO Image and Catalogue Archive operated by the United States Naval Observatory, Flagstaff Station (\url{http://www.nofs.navy.mil/data/fchpix/}) from \citet{Monet1998}.

Based on observations made with the NASA Galaxy Evolution Explorer. GALEX is operated for NASA by the California Institute of Technology under NASA contract NAS5-98034.

Some of the data presented in this paper were obtained from the Mikulski Archive for Space Telescopes (MAST). STScI is operated by the Association of Universities for Research in Astronomy, Inc., under NASA contract NAS5-26555. Support for MAST for non-HST data is provided by the NASA Office of Space Science via grant NNX13AC07G and by other grants and contracts.

The National Radio Astronomy Observatory is a facility of the National Science Foundation operated under cooperative agreement by Associated Universities, Inc.

This research has made use of data obtained from or software provided by the US National Virtual Observatory, which is sponsored by the National Science Foundation.

We acknowledge the use of public data from the Swift data archive.

This research has made use of data and/or services provided by the International Astronomical Union's Minor Planet Center.

This research uses services or data provided by the Science Data Archive at NOAO. NOAO is operated by the Association of Universities for Research in Astronomy (AURA), Inc. under a cooperative agreement with the National Science Foundation.

This work made use of the IPython package \citep{PER-GRA:2007}, Scikit-learn \citep{mckinney}, and Astropy, a community-developed core Python package for Astronomy \citep{2013A&A...558A..33A}.

The acknowledgements were (partially) compiled using the Astronomy Acknowledgement Generator.

\facility{CGRO, COBE, EUVE, Fermi, GALEX, GRANAT, HEASARC, IRSA, MAST, NRAO, OCA:IRIS, Planck, ROSAT, Sloan, Swift, VLA, WISE, WMAP, WSRT}

\software{Astropy \citep{2013A&A...558A..33A}; pyWWT \citep{robitaille_thomas_2017_1155830}}

\appendix
\label{sec:appendix}

\section{Curated Data}
\startlongtable
  \begin{deluxetable*}{p{2.5in}p{0.5in}p{3.5in}}
  \tablecolumns{3}
  \tablecaption{All Sky Surveys\label{table:allsky-data}}
  \tablehead{
  \colhead{Name (Bandpass)} &
  \colhead{Approximate} &
  \colhead{Notes, Credits} \\
  \colhead{} &
  \colhead{TOAST Resolution} &
  \colhead{} \\
  \colhead{} &
  \colhead{(arcsec/pixel)} &
  \colhead{}}
\startdata
 \cutinhead{Radio} \\
 VLSS: VLA Low-frequency Sky Survey  &  79.2 &  \citet{Cohen2007}\tablenotemark{a} \\
 VLA FIRST: Faint Images of the Radio Sky at Twenty-centimeters &  9.90 &  \citet{Becker1995} \\
 NVSS: NRAO VLA Sky Survey  &  19.8 &  \citet{Condon1998}\tablenotemark{a} \\
 SUMSS: Sydney University Molonglo Sky Survey  &  19.8 &  \citet{Bock1999} (SUMSS, Min=0.001, Max=10).\tablenotemark{a}  \\
 Westerbork Northern Sky Survey  &  9.90 &  WENSS Team. The WENSS project is a collaboration between the Netherlands Foundation for Research in Astronomy (NFRA/ASTRON) and the Leiden Observatory. (WENSS, Min=0.001, Max=10), who got it from the WENSS FTP site 1999-03-18. The original data was found using the Westerbork Synthesis Radio Telescope, which is operated by NFRA/ASTRON with support from the Netherlands Foundation for Scientific Research (NWO).\tablenotemark{a} \\
 Bonn 1420 MHz Survey  &  39.6 &  Max Planck Institute for Radio Astronomy, generated by Patricia Reich and Wolfgang Reich using with the Bonn Stockert 25m telescope. (1420MHz, Min=3, Max=35).\tablenotemark{a}  \\
 HI All-Sky Continuum Survey  &  158 &  Max Planck Institute for Radio Astronomy, generated by Glyn Haslam using data taken at Jodrell Bank, Effelsberg and Parkes telescopes. (408MHz, Min=10, Max=200).\tablenotemark{a}  \\
\cutinhead{Microwave} \\
 Planck
 \begin{itemize}
   \item[] CMB
   \item[] Dust \& Gas
   \item[] Thermal Dust
   \item[] Spinning Dust
   \item[] Molecular Gas (CO)
   \item[] Ionized Gas (free-free)
   \item[] Synchrotron (non-thermal)
   \item[] Lensing (Mass)
  \end{itemize} &  39.6 &  Planck is a European Space Agency mission, with significant participation from NASA. NASAs Planck Project Office is based at JPL. JPL contributed mission-enabling technology for both of Plancks science instruments. European, Canadian and U.S. Planck scientists work together to analyze the Planck data. \\
 WMAP
 \begin{itemize}
   \item[] ILC 5-Year Cosmic Microwave Background
   \item[] K Band, Linear, Non-Linear (23 GHz), and Polarization Map (50 uK)
   \item[] Ka Band, Linear, Non-Linear (33 GHz), and Polarization Map (35 uK)
   \item[] Q Band, Linear, Non-Linear (41 GHz), and Polarization Map (35 uK)
   \item[] V Band, Linear, Non-Linear (61 GHz), and Polarization Map (35 uK)
   \item[] W Band, Linear, Non-Linear (94 GHz), and Polarization Map (35 uK)
   \item[] WMAP QVW (Power and Linear)
 \end{itemize}
  &  158 &  NASA/WMAP Science Team  \\
 \cutinhead{Infrared} \\
 SFD Dust Map, 100 Micron  &  39.6 &  Data provided by two NASA satellites, the  Infrared Astronomy Satellite (IRAS) and the Cosmic Background Explorer (COBE). Processing by David J. Schlegel, Douglas P. Finkbeiner and Marc Davis, Princeton University and University of California, Berkeley. (SFD100m, Min=5, Max=5).\tablenotemark{a} \\
 WISE All Sky  &  19.8 &  NASA/JPL-Caltech/UCLA \\
 2MASS Two Micron All Sky Survey
 \begin{itemize}
   \item[] Imagery
   \item[] Synthetic catalog
   \end{itemize}  &
   \begin{itemize}
     \item[] 9.90
     \item[] 39.6
   \end{itemize}
  & Syntheic catalog image by Jina Suh from 2MASS catalog Copyright Microsoft 2007.  \\
 IRIS: Improved Reprocessing of IRAS Survey, 12, 25, 60, and 100 $\mu$m  &  39.6 &  Red is IRIS100, Green IRIS60, Blue IRIS12 (IRIS12: Min=0.5, Max=1; IRIS25: Min=1, Max=2; IRIS60: Min=1, Max=2; IRIS100: Min=4, Max=4)\tablenotemark{ab} \\
 COBE Diffuse Infrared Background Experiment (DIRBE)  &  39.6 &  NASA/COBE (COBE, Min=10, Max=5).\tablenotemark{a} \\
 COBE DIRBE Annual Average Map  &  39.6 &  NASA/COBE  (COBEAAM, Min=10, Max=5).\tablenotemark{a} \\
 COBE DIRBE Zodi-Subtracted Mission Average  &  39.6 &  NASA/COBE  (COBEZSMA, Min=10, Max=5).\tablenotemark{a} \\
\cutinhead{Optical} \\
 Hydrogen Alpha Full Sky Map &  39.6 &  Image Courtesy Douglas Finkbeiner. The full-sky H-alpha map (6' FWHM resolution) is a composite of the Virginia Tech Spectral line Survey (VTSS) in the north and the Southern H-Alpha Sky Survey Atlas (SHASSA) in the south. The Wisconsin H-Alpha Mapper (WHAM) survey provides a stable zero-point over 3/4 of the sky on a one degree scale.  \\
 SDSS: Sloan Digital Sky Survey &  0.30 &  Copyright SDSS \\
 Digitized Sky Survey &  0.62 &  Copyright DSS Consortium \\
 Tycho (Synthetic) &  79.2 &  NASA/Goddard Space Flight Center Scientific Visualization Studio \\
 USNOB: US Naval Observatory B 1.0 (Synthetic) &  39.6 &  \citet{Monet1998}. Rendering by Jina Suh (Microsoft). \\
\cutinhead{Ultraviolet} \\
 GALEX 2 Combined &  2.48 &  GR2/3 release\tablenotemark{ae} \\
 GALEX 2 Near-UV  &  2.48 &  GR2/3 release\tablenotemark{ae} \\
 GALEX 2 Far-UV   &  2.48 &  GR2/3 release\tablenotemark{ae}\\
 GALEX 4 Near-UV  &  2.48 &  Near UV band (1770-2730\AA) of the GR4 release\tablenotemark{ae}\\
 GALEX 4 Far-UV   &  2.48 &  Far UV band (1350-1780\AA) of the GR4 release\tablenotemark{ae} \\
 Extreme Ultraviolet Explorer (EUVE) &  79.2 &  Red is 555 \AA (euve555), Green 405 \AA, Red 83 \AA.\tablenotemark{ade} \\
 EUVE: 83 \AA &  79.2 & (euve83, Min=1, Max=150).\tablenotemark{ade} \\
 EUVE: 171 \AA &  79.2 &  (euve171, Min=4, Max=200).\tablenotemark{ade}\\
 EUVE: 405 \AA &  79.2 &  (euve405, Min=0.7, Max=70).\tablenotemark{ade}\\
 EUVE: 555 \AA &  79.2 &  (euve555, Min=5, Max=500).\tablenotemark{ade} \\
\cutinhead{X-ray} \\
 RASS: ROSAT All Sky Survey  &  19.8 &  This is a composite of three RASS3 surveys from the ROSAT Data Archive of the Max-Planck-Institut fur extraterrestrische Physik (MPE) at Garching, Germany. Red is soft band (RASS3sb), Green is broad band (RASS3bb), Blue is hard band (RASS3hb)\tablenotemark{a} \\
 ROSAT Hard Band Count Map  &  19.8 &  (RASS3hb, Min=0.2, Max=50).\tablenotemark{ac}\\
 ROSAT Soft Band Count Map  &  19.8 &  (RASS3sb, Min=0.2, Max=50).\tablenotemark{ac}\\
 ROSAT Broad Band Count Map  &  19.8 & (RASS3bb, Min=0.2, Max=50).\tablenotemark{ac}\\
 ROSAT Soft Band Intensity  &  19.8 &  (RASSSB, Min=0.2, Max=10).\tablenotemark{ac}\\
 ROSAT Hard Band Intensity  &  19.8 &   (RASSHB, Min=0.2, Max=10).\tablenotemark{ac} \\
 ROSAT PSPC summed pointed observations, 2 degree cutoff, intensity  &  9.90 &  Observational data from NASA Goddard Space Flight Center, (PSPC2int, Min=0.02, Max=0.01).\tablenotemark{a} \\
 Swift BAT All-Sky Survey: Significance 14-195 keV  &  79.2 &  NASA BAT Team. (BATSig, Min=1, Max=10).\tablenotemark{a} \\
 Swift BAT All-Sky Survey: Flux 14-195 keV  &  79.2 &  NASA BAT Team. (BATFlux, Min=1e-5, Max=1e-3).\tablenotemark{a} \\
 GRANAT/SIGMA Significance  &  79.2 &   (GRANAT\_SIGMA\_sig, Min=0.5, Max=50).\tablenotemark{af} \\
 GRANAT/SIGMA Flux  &  79.2 &  (GRANAT\_SIGMA\_flux, Min=1e-5, Max=0.01).\tablenotemark{af}  \\
\cutinhead{Gamma-ray} \\
 Fermi  &  158 &  NASA and the FERMI-LAT Team. \\
 Fermi Six Months  &  158 &  \multirow{4}{3.5in}{NASA and the FERMI-LAT Team.}\\
 Fermi Year Two  &  39.6 &   \\
 Fermi Year Three  &  79.2 &  \\
 Fermi LAT Year Eight  &  39.6 &  \\
 CGRO Compton Telescope: 3 channel data  &  79.2 &  CompTel Instrument Team. Maps generated by Andrew Strong, Max-Planck Institute for Extraterrestrial Physics, Garching, Germany. (comptel, Min=0.05, Max=0.1).\tablenotemark{a}  \\
 EGRET Soft  &  79.2 &  EGRET Instrument team, NASA Goddard Space Flight Center. (EGRETsoft, Min=0.1, Max=0.002).\tablenotemark{a} \\
 EGRET Hard  &  79.2 &  EGRET Instrument team, NASA Goddard Space Flight Center. (EGREThard, Min=0.02, Max=0.005).\tablenotemark{a} \\
\enddata
\tablenotetext{a}{TOAST-formatted mosaics were obtained using facilities of NASA’s SkyView Virtual Telescope.}
\tablenotetext{b}{Original IRAS data: NASA/JPL IPAC, IRIS Reprocessing: Canadian Institute for Theoretical Astrophysics/Institut d'Astrophysique Spatiale.}
\tablenotetext{c}{ROSAT Data Archive of the Max-Planck-Institut fur extraterrestrische Physik (MPE) at Garching, Germany.}
\tablenotetext{d}{Center for Extreme UV Astronomy, University of California at Berkeley.}
\tablenotetext{e}{Data archived at MAST/STScI}
\tablenotetext{f}{High Energy Astrophysics Department, Space Research Institute, Moscow, Russia; CEA, Centre d'Etudes de Saclay Orme des Merisiers, France; Centre d'Etude Spatiale des Rayonnements, Toulouse, France; Federation de Recherche Astroparticule et Cosmologie Universite de Paris, France.}
\end{deluxetable*}

\startlongtable
  \begin{deluxetable*}{p{2.5in}p{0.5in}p{3.5in}}
  \tablecolumns{3}
    \tablecaption{Panoramas\label{table:panorama-data}}
  \tablehead{
  \colhead{Name} &
  \colhead{Approximate} &
  \colhead{Notes, Credits} \\
  \colhead{} &
  \colhead{TOAST Resolution} &
  \colhead{} \\
  \colhead{} &
  \colhead{(arcsec/pixel)} &
  \colhead{}}
\startdata
Apollo 12: Landing Site &  19.8 &  Panorama taken by Apollo 12 Lunar Module Pilot Alan Bean in 1969 and assembled at the NASA Ames Research Center in 2007.  Images courtesy of NASA and the Lunar and Planetary Institute. \\
Apollo 17: Shorty Crater &  19.8 &  Panorama taken by Apollo 17 Commander Gene Cernan in 1972 and assembled at the NASA Ames Research Center in 2007.  Images courtesy of NASA and the Lunar and Planetary Institute. \\
Pathfinder: Improved MPF 360-degree Presidential Panorama &  79.1 &  NASA/JPL. This is the 1993 ``geometrically improved, color enhanced'' version of the 360-degree ``Gallery Pan'' at Ares Vallis,  the first contiguous, uniform panorama taken by the Imager for Mars (IMP) over the course of Sols 8, 9, and 10. \\
Pathfinder: ``Many Rovers'' &  79.1 &  \multirow{2}{3.5in}{Panoramas made available by Dr. Carol Stoker, NASA AMES. NASA/JPL/USGS.} \\
Pathfinder: Monster (stereo) &  79.1 &  \\
Pathfinder: Landing site from Sagan Memorial Station (stereo) &  39.6 &  NASA/JPL/USGS  \\
Spirit: McMurdo &  19.8 &  This 360-degree view, called the ``McMurdo'' panorama, comes from the panoramic camera (Pancam) on NASA's Mars Exploration Rover Spirit. From April through October 2006, Spirit has stayed on a small hill known as ``Low Ridge.'' There, the rover's solar panels are tilted toward the sun to maintain enough solar power for Spirit to keep making scientific observations throughout the winter on southern Mars. This view of the surroundings from Spirit's ``Winter Haven'' is presented in approximately true color. Image mosaicking: Kris Kapraro, Bob Deen, and the JPL/MIPL team.\tablenotemark{ac} \\
Spirit: McMurdo (false color) &  19.8 &  The Pancam began shooting component images of this panorama during Spirit's sol 814 (April 18, 2006) and completed the part shown here on sol 932 (Aug. 17, 2006). The panorama was acquired using all 13 of the Pancam's color filters. Image mosaicking: Kris Kapraro, Bob Deen, and the JPL/MIPL team.\tablenotemark{ac} \\
Spirit: McMurdo (color stereo) &  19.8 &  This 360-degree view comes from the panoramic camera (Pancam) on NASA's Mars Exploration Rover Spirit. From April through October 2006, Spirit has stayed on a small hill known as ``Low Ridge.'' \\
Spirit: McMurdo (left eye)  &  19.8 &  \multirow{7}{3.5in}{The Pancam on the Spirit Rover began shooting component images of this panorama during Spirit's sol 814 (April 18, 2006) and completed the part shown here on sol 932 (Aug. 17, 2006). The panorama was acquired using all 13 of the Pancam's color filters, using lossless compression for the red and blue stereo filters, and only modest levels of compression on the remaining filters. The overall panorama consists of 1,449 Pancam images.\tablenotemark{a}} \\
Spirit: McMurdo (right eye)  &  19.8 &   \\
  &   &   \\
  &   &   \\
  &   &   \\
  &   &   \\
  &   &   \\
Spirit: West Valley  &  19.8 & \multirow{3}{3.5in}{NASA'S Mars Exploration Rover Spirit captured this westward view from atop a low plateau where Sprit spent the closing months of 2007. NASA/JPL-Caltech/Cornell University.} \\
Spirit: West Valley (false color) &  19.8 &   \\
Spirit: West Valley (stereo) &  19.8 &   \\
\\tablebreak
Spirit: Descent from Husband Hill  &  19.8 &  \multirow{3}{3.5in}{Spirit acquired the 405 individual images that make up this 360-degree view of the surrounding terrain using five different filters on the panoramic camera. The rover took the images on Martian days, or sols, 672 to 677 (Nov. 23 to 28, 2005). \tablenotemark{abc}} \\
Spirit: Descent from Husband Hill (false color) &  19.8 &  \\
Spirit: Descent from Husband Hill (stereo) &  19.8 & \\
Spirit: Gibson Panorama at Home Plate (false color) &  19.8 &  \multirow{4}{3.5in}{NASA's Mars Exploration Rover Spirit acquired this high-resolution view of intricately layered exposures of rock while parked on the northwest edge of the bright, semi-circular feature known as ``Home Plate.'' The rover was perched at a 27-degree upward tilt while creating the panorama, resulting in the ``U'' shape of the mosaic. \tablenotemark{d} } \\
Spirit: Gibson Panorama at Home Plate &  39.6 &   \\
&   &   \\
&   &   \\
Spirit: Home Plate South &  19.8 &  This is the Spirit Pancam ``Home Plate South'' panorama, acquired on sols 1325 - 1332 (September 25 - October 2, 2007). This is an approximate true color rendering using Pancam's 753 nm, 535 nm, and 432 nm filters.\tablenotemark{a}   \\
Spirit: Home Plate South (false color) &  19.8 &  \multirow{2}{3.5in}{This is a false-color version of the Spirit Pancam ``Home Plate South'' panorama, acquired on sols 1325 - 1332 (September 25 - October 2, 2007).\tablenotemark{a} } \\
Spirit: Home Plate South (stereo) &  19.8 &  \\
Spirit: Husband Hill Summit &  39.6 &  The panoramic camera on NASA's Mars Exploration Rover Spirit took the hundreds of images combined into this 360-degree view, the ``Husband Hill Summit'' panorama. The images were acquired on Spirit's sols 583 to 586 (Aug. 24 to 27, 2005), shortly after the rover reached the crest of ``Husband Hill'' inside Mars' Gusev Crater.  \\
Spirit: Santa Anita  &  19.8 &  This color mosaic was taken on May 21, 25 and 26, 2004, by the panoramic camera on NASA's Mars Exploration Rover Spirit was acquired from a position roughly three-fourths the way between ``Bonneville Crater'' and the base of the ``Columbia Hills''.\tablenotemark{a}  \\
Spirit: Independence  &  19.8 &  The 108 images used to make this panorama were obtained on sols 536-543 (July 6 to 13, 2005) from a position in the Columbia Hills near the summit of Husband Hill. \\
Spirit: Everest  &  19.8 & It took Spirit three days, sols 620 to 622 (Oct. 1 to Oct. 3, 2005), to acquire all the 81 images combined into this mosaic, looking outward in every direction from the true summit of Husband Hill. Sky fixed by Jim St George.\tablenotemark{abc} \\
Spirit: Everest (stereo) &  19.8 & Composed of 81 images acquired on sols 620 to 622 (October 1 to 3, 2005) from a position in the Columbia Hills at the true summit of Husband Hill.\tablenotemark{ab}  \\
Spirit Mission Success &  39.6 & This panorama was acquired in eight parts called octants, over the course of three different sols. Because the octants were taken at different times of day and across different days when the dust abundance was changing, there were large brightness and color seams between the octants in the assembled mosaic.\tablenotemark{a}  \\
Spirit: Thanksgiving  &  19.8 &  \multirow{4}{3.5in}{This is the Spirit Pancam ``Thanksgiving'' panorama, acquired on sols 318 to 325 (Nov. 24 to Dec. 2, 2004) from a position along the flank of Husband Hill, which is the peak just left of the center of this mosaic, just east of the West Spur of the Columbia Hills.\tablenotemark{abc}} \\
Spirit: Thanksgiving (stereo) &  79.1 &  \\
 & & \\
 & & \\
 \\tablebreak
Spirit: Cahokia  &  19.8 &  \multirow{6}{3.5in}{The approximate true-color image, nicknamed the ``Cahokia panorama'' after the Native American archaeological site near St. Louis, was acquired between sols 213 to 223 (Aug. 9 to 19, 2004). The panorama consists of 470 images acquired through six panoramic camera filters (750 to 480 nanometers). Stereo images: JPL/MIPL (Bob Deen).\tablenotemark{a}} \\
Spirit: Cahokia (stereo, unadjusted) &  19.8 &  \\
Spirit: Cahokia (stereo, tilt-adjusted) &  158 & \\
 & & \\
 & & \\
 & & \\
Spirit: Larry's Lookout  &  19.8 &   \multirow{4}{3.5in}{This is the Spirit Pancam ``Lookout'' panorama, acquired on sols 410 to 413 (Feb. 27 to Mar. 2, 2005) from a position known informally as ``Larry's Lookout'' along the drive up Husband Hill.\tablenotemark{abc}} \\
Spirit: Larry's Lookout (stereo) &  19.8 &  \\
 & & \\
  & & \\
Spirit: Whale of a Panorama  &  19.8 &  Spirit produced this 220 degree image mosaic two-thirds of the way to the summit of Husband Hill from images collected from sol 497 to 500 (May 27 through May 30, 2005).\tablenotemark{a}   \\
Spirit: From The Summit &  39.6 &  This approximate true-color 240 degree panorama was taken by NASA's Spirit rover from the top of ``Husband Hill''  in the ``Columbia Hills'' of Gusev Crater. The mosaic is made up of images taken by the rover's panoramic camera over a period of three days (sols 583 to 585, or August 24 to 26, 2005).  \\
Spirit: Paige Panorama  &  39.6 &  \multirow{2}{3.5in}{This 230-degree panorama was composed from 72 images from the Pancam on Spirit on Feb 19 2006.\tablenotemark{d}} \\
Spirit: Paige Panorama (false color) &  39.6 &  \\
Spirit: Legacy  &  19.8 & The 78 images used to make this were acquired on sols 59 to 61 (March 3 to 5, 2004) from a position about halfway between the landing site and the rim of Bonneville crater, within the transition from the relatively smooth plains to the more rocky and rugged ejecta blanket of Bonneville.\tablenotemark{abc} \\
Spirit: Bonneville &  19.8 &  The rim and interior of a crater nicknamed ``Bonneville'' dominate this 180-degree, false-color mosaic. Spirit recorded this view on the rover's 68th sol, March 12, 2004, one sol after reaching this location. The rover remained here in part to get this very high-resolution, color mosaic, from which scientists can gain insight about the depth of the surface material at Bonneville and make future observation plans. On sol 71, Spirit was instructed to drive approximately 15 meters (49 feet) along the crater rim to a new vantage point.\tablenotemark{a}  \\
Opportunity: Erebus  &  19.8 &  \multirow{4}{3.5in}{This panorama was made from 635 images acquired on sols 652 to 663 (Nov. 23 to Dec. 5, 2005), as NASA's Mars Exploration Rover Opportunity was exploring sand dunes and outcrop rocks in Meridiani Planum. Image mosaicking: JPL/MIPL Team (Bob Deen, Oleg Pariser, Jeffrey Hall)\tablenotemark{ac}.}  \\
Opportunity: Erebus (Stereo) &  19.8 &  \\
 & & \\
 & & \\
 \\tablebreak
Opportunity: Lyell  &  39.6 &  \multirow{9}{3.5in}{This view combines many images taken by Opportunity's panoramic camera (Pancam) from the 1,332nd through 1,379th Martian days, or sols, of the mission (Oct. 23 to Dec. 11, 2007). The main body of the crater appears in the upper right of this stereo panorama, with the far side of the crater lying about 800 meters (half a mile) away. Bracketing that part of the view are two promontories on the crater's rim at either side of Duck Bay. They are ``Cape Verde,'' about 6 meters (20 feet) tall, on the left, and ``Cabo Frio,'' about 15 meters (50 feet) tall, on the right. The rest of the image, other than sky and portions of the rover, is ground within Duck Bay. Image mosaicking: Jim Bell and Jonathan Joseph.\tablenotemark{ac} }   \\
Opportunity: Lyell (Stereo) &  19.8 &  \\
 & & \\
 & & \\
 & & \\
 & & \\
 & & \\
 & & \\
 & & \\
Opportunity: Rub al Khali  &  19.8 &  Taken on the plains of Meridiani during the period from the rover's 456th to 464th sols on Mars (May 6 to May 14, 2005). Opportunity was about 2 kilometers (1.2 miles) south of ``Endurance Crater'' at a place known informally as ``Purgatory Dune''.\tablenotemark{a}  \\
\\tablebreak
Opportunity: Endurance South  &  19.8 &  \multirow{3}{3.5in}{This 360-degree panorama shows ``Endurance Crater'' and the surrounding plains of Meridiani Planum. This is the second large panoramic camera mosaic of Endurance, and was obtained from a high point near the crater's south rim.\tablenotemark{a}} \\
Opportunity: Endurance South (false color) &  19.8 &  \\
 & & \\
Opportunity Mission Succcess &  19.8 &  This expansive view of the martian real estate surrounding the Mars Exploration Rover Opportunity is the first 360 degree, high-resolution color image taken by the rover's panoramic camera. The airbag marks, or footprints, seen in the soil trace the route by which Opportunity rolled to its final resting spot inside a small crater at Meridiani Planum, Mars. The exposed rock outcropping is a future target for further examination. This image mosaic consists of 225 individual frames.  \\
Opportunity: Burns Cliff &  39.6 &  This view of ``Burns Cliff'' after driving right to the base of this southeastern portion of the inner wall of ``Endurance Crater.'' The view combines 46 images taken by Opportunity's panoramic camera between the rover's 287th and 294th martian days (Nov. 13 to 20, 2004).\tablenotemark{a}   \\
Opportunity: Payson Panorama  &  79.1 &  The panoramic camera aboard NASA's Mars Exploration Rover Opportunity acquired this panorama of the Payson outcrop on the western edge of Erebus Crater during Opportunity's sol 744 (Feb. 26, 2006). USGS\tablenotemark{a}.  \\
Opportunity: Beagle Crater  &  19.8 &  \multirow{6}{3.5in}{Opportunity took the mosaic of images that make up this 360-degree view of the rover's surroundings with the panoramic camera on the rover's 901st through 904th sols, or Martian days (Aug. 6 through Aug. 9, 2006), of exploration. This is an approximate true-color image combining exposures taken through the panoramic camera's 753-nanometer, 535-namometer, and 432-nanometer filters. Image mosaicking: Cornell Pancam team. NASA/JPL/Cornell/UNM\tablenotemark{c}} \\
Opportunity: Beagle Crater (false color) &  19.8 &   \\
 & & \\
 & & \\
 & & \\
 & & \\
Opportunity: Lion King &  19.8 &  This shows ``Eagle Crater'' and the surrounding plains of Meridiani Planum. It was stitched from 558 images obtained on sols 58 and 60 by the Mars Exploration Rover Opportunity's panoramic camera.\tablenotemark{a}  \\
Phoenix: Landing Site &  39.6 &  This view combines more than 400 images taken during the first several weeks after NASA's Phoenix Mars Lander arrived on an arctic plain at 68.22 degrees north latitude, 234.25 degrees east longitude on Mars. NASA/JPL-Caltech/University Arizona/Texas A \& M University.   \\
\enddata
\tablenotetext{a}{NASA/JPL/Cornell.}
\tablenotetext{b}{Image mosaicking: JPL/MIPL (Bob Deen)}
\tablenotetext{c}{Calibration and color rendering: Cornell Calibration Crew and the Pancam team (Jim Bell)}
\tablenotetext{d}{NASA/JPL-Caltech/USGS/Cornell.}
\end{deluxetable*}

\startlongtable
  \begin{deluxetable*}{p{2.in}p{0.5in}p{3.5in}}
  \tablecolumns{3}
    \tablecaption{Planet Maps\label{table:planet-data}}
  \tablehead{
  \colhead{Name} &
  \colhead{Approximate} &
  \colhead{Notes, Credits} \\
  \colhead{} &
  \colhead{TOAST Resolution} &
  \colhead{} \\
  \colhead{} &
  \colhead{(arcsec/pixel)} &
  \colhead{}}
\startdata
Bing Maps Aerial &  0.002 &  \multirow{3}{3.5in}{Bing Maps Copyright Microsoft Corp. and Suppliers} \\
Bing Maps Hybrid &  0.002 &   \\
Bing Maps Streets &  0.002 &   \\
Earth at Night 2012 &  4.94 &  NASA Earth Observatory/NOAA NGDC \\
Earth at Night &  19.8 &  Data courtesy Marc Imhoff of NASA GSFC and Christopher Elvidge of NOAA NGDC. Image by Craig Mayhew and Robert Simmon, NASA GSFC. \\
Blue Marble January 2004 &  19.8 &  Blue Marble Next Generation, January 2004. Visible Earth Project, NASA. \\
Blue Marble July 2004 &  19.8 &  Blue Marble Next Generation, July 2004. Visible Earth Project, NASA. \\
Open Street Maps &  0.010 &  Map data Copyright OpenStreetMap contributors, CC BY-SA \\
Open Street Maps Bike Map &  0.010 &  Map data Copyright OpenStreetMap contributors, CC BY-SA \\
Moon &  2.47 &  Image Courtesy NASA/JPL \\
Mercury &  9.89 &   NASA/Johns Hopkins University Applied Physics Laboratory/Carnegie Institution of Washington \\
Venus &  79.1 &  Image Courtesy NASA/JPL/Space Science Institute \\
Mars Visible Imagery &  9.89 &  NASA/USGS/Malin Space Science Systems/JPL \\
Jupiter &  158 &  Image Courtesy NASA/JPL/Space Science Institute \\
Phobos (Mars) &  316 &  Viking images of the moon Phobos.\tablenotemark{a}  \\
Deimos (Mars) &  316 &  Viking images of the moon Deimos.\tablenotemark{a}  \\
Io (Jupiter) &  79.1 &  \multirow{4}{3.5in}{Images Courtesy NASA/JPL/Space Science Institute.}  \\
Europa (Jupiter) &  79.1 &   \\
Ganymede (Jupiter) &  79.1 & \\
Callisto (Jupiter) &  79.1 &  \\
Mimas (Saturn) &  316 &   Approximately half the surface of moon Mimas\tablenotemark{ab} \\
Enceladus (Saturn) &  316 &   Approximately half the surface of moon Enceladus\tablenotemark{ab} \\
Tethys (Saturn) &  316 &   Most of the surface of the moon Tethys\tablenotemark{ab} \\
Dione (Saturn) &  316 &   Most of the surface of the moon Dione\tablenotemark{ab}  \\
Rhea (Saturn) &  316 &   The moon Rhea\tablenotemark{ab} \\
Iapetus (Saturn) &  316 &   Approximately half of the surface of the moon Iapetus\tablenotemark{ab} \\
Ariel (Uranus) &  316 &   Most of the Southern Hemisphere of the moon Ariel\tablenotemark{ab} \\
Umbriel (Uranus) &  316 &   Most of the Southern Hemisphere of the moon Umbriel\tablenotemark{ab} \\
Titania (Uranus) &  316 &   Most of the Southern Hemisphere of the moon Titania\tablenotemark{ab} \\
Oberon (Uranus) &  316 &   Most of the Southern Hemisphere of the moon Oberon\tablenotemark{ab} \\
Miranda (Uranus) &  316 &   Most of the Southern Hemisphere of the moon Miranda\tablenotemark{ab} \\
\enddata
\tablenotetext{a}{Caltech/JPL/USGS.}
\tablenotetext{b}{From Voyager mission photographs.}
\end{deluxetable*}

\bibliography{references}

\end{document}